\documentclass{preprint}

\graphicspath{{./figs/}}

\newcommand\changes[1]{\textcolor{black}{#1}}

\titlehead{Research}

\begin{document}

\title{Phenotype-structured nonlocal aggregation models I: signal production and response}

\author{
Thomas Jun Jewell$^{1}$, Andrew L. Krause$^{2}$, Philip K. Maini$^{1}$, and Eamonn A. Gaffney$^{1}$}

\address{$^{1}$Wolfson Centre for Mathematical Biology, Mathematical Institute, University of Oxford, Andrew Wiles Building, Radcliffe Observatory Quarter, Woodstock Road, Oxford, OX2 6GG, United Kingdom\\
$^{2}$Mathematical Sciences Department, Durham University, Upper Mountjoy Campus, Stockton Rd, Durham, DH1 3LE, United Kingdom}

\subject{mathematical modelling, applied mathematics}


\corres{Thomas Jun Jewell\\
\email{jewell@maths.ox.ac.uk}}

\begin{abstract}
\changes{Individuals within a cellular or ecological population can exhibit substantial variation in their traits or phenotype. Such heterogeneity can fundamentally alter population-level behaviour, including the ability to aggregate and form spatial patterns.} We investigate these effects by extending nonlocal aggregation models, widely used to model cell--cell adhesion and collective animal movement, to incorporate phenotype structure. We characterise how heterogeneity and switching between phenotypes affect whether patterns form, how phenotypes are distributed within them, and whether they are stationary or exhibit spatio-temporal oscillations. We show that pattern formation is promoted when phenotypes that produce stronger interaction signals also respond more strongly to them, whereas specialisation towards either signal production or response inhibits pattern formation. Increasing rates of switching between phenotypes dampen these effects. Heterogeneity can also enable weakly interacting, and even repulsive, individuals to pattern as part of the collective. Furthermore, movement in phenotype and physical space can interact to restructure patterns and generate spatio-temporal oscillations. \changes{A further key message of this study is the value of considering both discrete $N$-phenotype and continuous phenotype formulations. We develop both in parallel, demonstrating formal correspondences, but also advantages to each framework in terms of mathematical analyses and interpretability.} \absbreak
\end{abstract}

\maketitle


\newpage
\section{Introduction}

Phenotypic heterogeneity is ubiquitous in cellular and ecological populations, where individuals can vary substantially in their morphology, behaviour, physiology and interactions with others. \changes{In ecology, within-population trait variation can influence population dynamics and species interactions \cite{bolnick2011_ecology_trait_variation}, and has long been studied using structured-population models, particularly those dividing populations into discrete or continuous classes according to age, size, or behavioural strategy \cite{tuljapurkar2012_ecology_structured_population_book,cushing1998_structured_population_book}.} Cellular populations similarly exhibit heterogeneity in traits such as signalling, motility, and proliferation, with particularly important consequences in cancer \cite{lorenzi2025phenotype_tutorial,carter2021heterogeneity_phenotype,ackermann2015heterogeneity_phenotype,keller2019heterogeneity_phenotype}. Moreover, \changes{traits} need not fall into a small number of discrete classes. For example, epithelial--mesenchymal plasticity can generate intermediate states along a phenotypic spectrum \cite{hari_jolly2022EMT}, motivating models in which populations are structured by a continuous phenotype variable \cite{lorenzi2025phenotype_tutorial,yang2020_continuum_phenotype}. \changes{We will explore the impact of such phenotypic variation on the ability of populations to aggregate and form spatial niches, or patterns.}

One widely used framework for studying how interacting populations organise in space is provided by nonlocal aggregation, or nonlocal advection--diffusion, models \cite{painter2024_nonlocal_review}. Initially developed to study animal swarms, where individuals respond to neighbours over finite distances \cite{mogilner_edesteinkeshet_1999nonlocal_swarm}, these models were subsequently adopted to describe cell--cell adhesion \cite{armstrong2006continuum}. Here, interactions naturally occur across the cell membrane and can extend further through cellular protrusions such as pseudopodia \cite{painter2015nonlocal}. Nonlocal aggregation models have since become an established framework for cell--cell adhesion and other forms of collective cell movement, as well as spatial organisation in ecological populations, for example driven by scent marking and memory \cite{painter2024_nonlocal_review,potts2019_memory}.

In a standard nonlocal aggregation model
\cite{armstrong2006continuum,painter2015nonlocal,giunta2023positivity_existence,jewell2023_dimension,jewell2025chase,yurij_jose_2026_bifurcation_two_species_nonlocal,painter2024_nonlocal_review},
the evolution of a single population with density $u(t,\mathbf{x})$ on a $d$-dimensional domain $\Omega$ is described by a combination of random diffusion and directed motion induced by nonlocal interactions:
\begin{equation}
    \frac{\partial u(t,\mathbf{x})}{\partial t}
    =
    \nabla^2u
    -
    \nabla\cdot
    \left(
    u\,p(u)
    \nabla
    \int_{\Omega}
    K(s)\,
    u(t,\mathbf{x}+\mathbf{s})
    \,\mathrm{d}^d\mathbf{s}
    \right).
\label{eq:single_species_model}
\end{equation}
The interaction kernel $K(s)$ determines the strength of interactions between individuals separated by a distance $s=|\mathbf{s}|$, while the saturation function $p(u)$ can account for reduced motility due to crowding or volume-filling effects. In this formulation, however, all individuals are effectively interchangeable. \changes{Many studies extend this by considering two distinct species, with a few briefly exploring up to six \cite{potts2019_memory,giunta2025phylogeny}. In this work, we will extend the framework to populations with arbitrarily many phenotypes, or even a continuum of phenotypic states.}

\changes{Pattern formation in phenotype-structured models remains relatively underexplored, although early examples can be found in the structured-population literature \cite{Gourley2008_pattern_in_structured }. More recently, interest in this area has grown, with studies considering phenotype-structured Keller--Segel chemotaxis models \cite{lorenzi_painter2025phenotype_chemotaxis} and Shigesada--Kawasaki--Teramoto competition models \cite{cusseddu_lorenzi_2026phenotype_SKT}.}

Despite this increasing interest, the recent review by Lorenzi et al.~\cite{lorenzi2025phenotype_tutorial} identifies only two examples of phenotype-structured nonlocal aggregation models. Engwer et al.~\cite{engwer2017_nonlocal_phenotype_acid} introduced a phenotype-structured model of acid-mediated tumour invasion, establishing existence of solutions for a biologically detailed system incorporating adhesion alongside additional processes \changes{specific to their application}. More recently, Lorenzi et al.~\cite{lorenzi2024_nonlocal_kinetic_phenotype} derived a phenotype-structured nonlocal migration model from kinetic theory and investigated its behaviour in the context of a cell migration assay. Existence theory has also been developed for nonlocal aggregation models with an arbitrary finite number of interacting species \cite{giunta2023positivity_existence}.

Phenotype-structured aggregation models also introduce substantial analytical challenges. In a discrete description, the number of interacting phenotypes may be arbitrarily large, while a continuous description is doubly nonlocal, with nonlocal interactions in both physical and phenotype space \cite{lorenzi2025phenotype_tutorial}.  Furthermore, stability analysis of pattern formation in phenotype-structured models is considerably more involved, since the relevant eigenfunctions are no longer simply those of the Laplacian and must instead be determined as part of the corresponding eigenvalue problem. Nevertheless, as we show here, considerable analytical progress remains possible.

In this work, we study the influence of phenotypic heterogeneity on how populations aggregate in space. We do so by extending nonlocal aggregation models to incorporate phenotype structure, focusing on interactions in which phenotype determines how strongly individuals produce interaction signals and how strongly they respond to them. We aim to characterise how these interactions influence the conditions for pattern formation, how the population is distributed across phenotype space within the resulting patterns, and whether these patterns are stationary or oscillate in time.

\changes{This study forms Part I of a two-part investigation into how phenotypic heterogeneity within a population affects pattern formation. While here we consider interactions that depend on the absolute phenotypes of individuals, in Part II \cite{jewell2026phenotype_part2} we will instead consider interactions that depend on relative phenotype, where interaction strength is determined by how similar two individuals are, as measured by their distance in phenotype space. Although both studies share much of the same mathematical framework, their biological interpretations and resulting behaviour differ substantially.}

We develop both discrete and continuous phenotype formulations of the model in parallel, establishing the relationship between them. \changes{Whilst continuous phenotype formulations are prominent in recent cell and cancer modelling literature \cite{lorenzi2025phenotype_tutorial, lorenzi_painter2025phenotype_chemotaxis, cusseddu_lorenzi_2026phenotype_SKT, engwer2017_nonlocal_phenotype_acid},} we show that an $N$-phenotype discrete formulation is similarly useful and requires fewer assumptions. Rather than favouring either description, we demonstrate the benefits of considering both: doing so reveals complementary mathematical structures and allows us to use whichever formulation is more natural for a given aspect of the problem.

We begin in Section~\ref{sec:model_formulation} by defining how phenotype affects interactions and formulating the discrete and continuous models, deriving one from the other. In Section~\ref{sec:linear_stability}, we develop the linear stability analysis of pattern formation in these models and establish the relationship between the two formulations. Sections~\ref{sec:slow_switching} and \ref{sec:fast_switching} analyse the limits of slow and fast phenotypic switching, respectively, before Section~\ref{sec:generic_switching} considers generic switching rates and the emergence of spatio-temporal behaviour. Finally, we discuss the broader implications and future directions in Section~\ref{sec:discussion}. Readers primarily interested in the biological insights can focus on the ``Biological insight'' subsections of Sections~\ref{sec:slow_switching}, \ref{sec:fast_switching} and \ref{sec:generic_switching}, together with the spatio-temporal results in Section~\ref{sec:spatio-temporal_oscillations} and the final Discussion.

\section{Phenotype-structured nonlocal aggregation models}
\label{sec:model_formulation}

To incorporate phenotypic heterogeneity into the nonlocal aggregation model \eqref{eq:single_species_model}, we consider two complementary formulations. The first consists of $N$ discrete phenotypes (or species), whilst the second replaces the discrete phenotype index with a continuous phenotype variable. Studying these formulations side-by-side allows us to establish the relationship between them, compare the mathematical structures that arise in each, and assess the complementary insights that each provides into pattern formation.

Following the standard approach in the nonlocal aggregation literature\,\cite{armstrong2006continuum,painter2015nonlocal,giunta2023positivity_existence,jewell2023_dimension,jewell2025chase,yurij_jose_2026_bifurcation_two_species_nonlocal,painter2024_nonlocal_review}, we assume periodic boundary conditions throughout, so that the physical domain is the torus, $\mathbf{x}\in\Omega=\mathbb{T}^d$. This allows us to focus on behaviour arising from interactions in the bulk, rather than from boundary effects, while avoiding the need to prescribe how nonlocal interactions behave across a boundary. We choose the domain sufficiently large to minimise spurious interaction due to wrap-around effects.

\subsection{Discrete phenotype formulation}

The discrete formulation is a generalisation of \eqref{eq:single_species_model} in which $N$ discrete phenotypes (or species), with densities $u_i(t,\mathbf{x})$, evolve according to \changes{the non-dimensional equations}
\begin{equation}
    \frac{\partial u_i(t,\mathbf{x})}{\partial t}
    =
    \nabla_{\mathbf{x}}^2u_i
    -
    \nabla_{\mathbf{x}}\cdot
    \left(
    u_i\,p(\overline{u})
    \nabla_{\mathbf{x}}
    \int_{\Omega}
    \sum_{j=1}^N
    W_{ij}(s)\,
    u_j(t,\mathbf{x}+\mathbf{s})
    \,\mathrm{d}^d\mathbf{s}
    \right)
    +
    \tilde{\beta}\sum_{j=1}^{N}L_{ij}u_j,
\label{eq:discrete_model}
\end{equation}
for $i=1,\ldots,N$. Here, $W_{ij}(s)$ denotes the interaction kernel governing the movement of phenotype $i$ induced by phenotype $j$ when they are separated by a distance $s=|\mathbf{s}|$. We also allow for phenotypic switching through the last term in \eqref{eq:discrete_model}, where $\tilde{\beta}\geq 0$ is the discrete switching rate and $L_{ij}$ is the graph Laplacian on the one-dimensional phenotype lattice satisfying no-flux boundary conditions, defined by
\begin{equation}
\sum_{j=1}^{N}L_{ij}u_j=
\begin{cases}
u_{i-1}-2u_i+u_{i+1}, & i=2,\ldots,N-1,\\[4pt]
-u_1+u_2, & i=1,\\[4pt]
u_{N-1}-u_N, & i=N.
\end{cases}
\label{eq:graph_laplacian}
\end{equation}
This models uniform stochastic switching or mutation between neighbouring phenotypes and is a common modelling choice in phenotype-structured systems
\cite{lorenzi2025phenotype_tutorial,lorenzi_painter2025phenotype_chemotaxis,chisholm_lorenzi2015cancer_phenotype}.

The saturation function $p(\overline{u})\geq 0$ \changes{depends on the local population density, $\overline{u}(t,\mathbf{x})$, averaged over phenotype,} and models reduced motility arising from crowding or volume-filling effects. A common choice is
\[
p(\overline{u})
=1-\overline{u}\equiv
1-\frac{1}{N}\sum_{j=1}^N u_j(t,\mathbf{x}),
\]
\changes{which enforces $\overline{u}(t,\mathbf{x})\leq 1$}. 

Importantly, this discrete formulation allows for an arbitrary number of phenotypes, with $N$ potentially small or large, and requires no assumptions about continuity or smoothness with respect to phenotype.

\subsection{Continuous phenotype formulation}

In the continuous formulation, the discrete phenotype index is replaced by a continuous phenotype variable $\theta\in V$, where $V\subset\mathbb{R}$ is referred to as the \textit{phenotype domain} or \textit{phenotype space}. The geometry of \(V\), together with the phenotypic dependence of the interaction kernel, encode our assumptions about how the population is structured by phenotype. The population density $u(t,\mathbf{x},\theta)$ depends on phenotype as well as (physical) space and time, and evolves according to \changes{the non-dimensional equations}
\begin{equation}
    \frac{\partial u(t,\mathbf{x},\theta)}{\partial t}
    =
    \nabla_{\mathbf{x}}^2u
    -
    \nabla_{\mathbf{x}}\cdot
    \left(
    u\,p(\overline{u})
    \nabla_{\mathbf{x}}
    \int_{\Omega}
    \int_V
    W(s,\theta,\theta')
    u(t,\mathbf{x}+\mathbf{s},\theta')
    \,\mathrm{d}\theta'
    \,\mathrm{d}^d\mathbf{s}
    \right)
    +
    \beta\frac{\partial^2u}{\partial\theta^2}.
\label{eq:continuum_model}
\end{equation}
Throughout, subscripts of $\mathbf{x}$ on differential operators indicate differentiation with respect to physical space only.

Here, the interaction kernel $W(s,\theta,\theta')$ governs the movement of individuals with phenotype $\theta$ induced by individuals with phenotype $\theta'$, when they are separated by a \changes{physical} distance $s=|\mathbf{s}|$. We allow for phenotypic switching through the last term in \eqref{eq:continuum_model}, where $\beta\ge0$ is the continuum switching rate. This represents linear diffusion in phenotype space and is the continuum analogue of the graph Laplacian in \eqref{eq:graph_laplacian}. Throughout, we impose no-flux conditions at the boundaries of the phenotype domain,
\begin{equation}
\frac{\partial u}{\partial\theta}=0,
\qquad
(t,\mathbf{x},\theta)\in(0,\infty)\times\Omega\times\partial V.
\label{eq:phenotype_BC}
\end{equation}

Analogously to the discrete formulation, the saturation function depends on the total local population density, with a common choice being
\[
p(\overline{u})
=1-\overline{u}\equiv
1-\frac{1}{|V|}
\int_V
u(t,\mathbf{x},\theta')
\,\mathrm{d}\theta'\geq 0.
\]

\changes{In both formulations, in order to investigate pattern formation, we typically consider initial conditions of a homogeneous state, $U$, with a small random perturbation that is heterogeneous across physical space and phenotypes.}

Although phenotype-structured models are \changes{often} formulated using a continuous phenotype variable \cite{lorenzi2025phenotype_tutorial}, we develop the discrete and continuous formulations in parallel throughout this work. This enables a direct comparison of their mathematical structures, as well as the analytical tractability and insight that each offers.

\subsection{Derivation of the continuous formulation from the discrete formulation}

The continuous phenotype model \eqref{eq:continuum_model} arises naturally as a formal continuum limit of the discrete phenotype model \eqref{eq:discrete_model} as the number of phenotypes tends to infinity. To construct this limit, we consider discrete models whose phenotypes increasingly finely sample a continuous phenotype space, and assume that the discrete phenotype densities and interaction kernels approach well-defined continuous functions. These additional assumptions are required only for the continuum limit; the general discrete formulation \eqref{eq:discrete_model} has broader scope, allowing an arbitrary number of phenotypes without requiring their properties to vary continuously or smoothly. The specific regularity conditions needed to justify each limiting step are stated below.

We assume that the phenotype domain $V$ is partitioned into $N$ intervals of equal width
\begin{equation}
    \Delta\theta=\frac{|V|}{N},
\end{equation}
with phenotype $i$ associated with the midpoint $\theta_i$ of the $i$th interval. The discrete phenotype densities are identified with samples of a phenotype density $u(t,\mathbf{x},\theta)$, where
\begin{equation}
    u_i(t,\mathbf{x})
    =
    u(t,\mathbf{x},\theta_i).
\end{equation}
Similarly, the discrete interaction kernels are identified with samples of an interaction kernel $W(s,\theta,\theta')$ according to
\begin{equation}
    W_{ij}(s)
    =
    W(s,\theta_i,\theta_j)\,\Delta\theta.
\label{eq:W_ij_W_theta_theta'_correspondence}
\end{equation}
The factor of $\Delta\theta$ reflects that each discrete phenotype represents an interval of phenotype space. Thus, $W_{ij}(s)$ represents the interaction contributed by individuals within the corresponding phenotype interval and consequently scales as $\mathcal{O}(N^{-1})$.

With these definitions, the interaction term in \eqref{eq:discrete_model} becomes
\begin{equation}
    \sum_{j=1}^N
    W_{ij}(s)\,
    u_j(t,\mathbf{x}+\mathbf{s})
    =
    \sum_{j=1}^N
    W(s,\theta_i,\theta_j)\,
    u(t,\mathbf{x}+\mathbf{s},\theta_j)\,
    \Delta\theta.
\end{equation}
Provided $W(s,\theta,\theta')\,
u(t,\mathbf{x}+\mathbf{s},\theta')$ is continuous in $\theta'$, this is a midpoint Riemann sum and, in the limit $N\changes{=\frac{|V|}{\Delta\theta}}\rightarrow\infty$, converges to
\begin{equation}
    \int_V
    W(s,\theta,\theta')\,
    u(t,\mathbf{x}+\mathbf{s},\theta')
    \,\mathrm{d}\theta',
\end{equation}
where $\theta_i\to\theta$ in the continuum limit.

Finally, we relate the discrete and continuous switching rates by
\begin{equation}
    \tilde{\beta}
    =
    \frac{\beta}{(\Delta\theta)^2}
    \sim \mathcal{O}(N^2),
\end{equation}
where $\beta\sim\mathcal{O}(1)$ is the continuous switching rate. Assuming $u$ is twice continuously differentiable with respect to $\theta$, the graph Laplacian in \eqref{eq:graph_laplacian} is the standard centred finite-difference approximation to the phenotype Laplacian,
\begin{equation}
    \tilde{\beta}\sum_{j=1}^{N}L_{ij}u_j
    =
    \beta
    \frac{u_{i-1}-2u_i+u_{i+1}}{(\Delta\theta)^2}
    \longrightarrow
    \beta\frac{\partial^2u}{\partial\theta^2},
\end{equation}
as $N\rightarrow\infty$. The discrete no-flux boundary conditions similarly converge to the continuous boundary condition \eqref{eq:phenotype_BC}.

Substituting these limits into \eqref{eq:discrete_model} yields the continuous phenotype model \eqref{eq:continuum_model}. The same construction extends naturally to multidimensional phenotype spaces.

\subsection{Separable spatial and phenotypic kernels}

The interaction kernels $W_{ij}(s)$ and $W(s,\theta,\theta')$ determine how interactions depend on physical separation and phenotype. In the most general case, the strength, direction, and spatial scale may all vary with the phenotypes of both interacting individuals. Within this broad framework, we focus on several biologically relevant interaction structures that allow us to isolate and understand how different aspects of phenotypic heterogeneity influence aggregation and pattern formation.

We assume that the spatial and phenotypic dependence of the interaction kernel are separable. Hence, in the discrete formulation,
\begin{equation}
    W_{ij}(s)=\tilde{\mu}\, A_{ij}K(s),
\label{eq:discrete_spatial_phenotype_separable}
\end{equation}
where $K(s)$ is a common spatial interaction kernel and $A_{ij}$ are the entries of the phenotype interaction matrix. Correspondingly, in the continuous formulation,
\begin{equation}
    W(s,\theta,\theta')=\mu\, A(\theta,\theta')K(s),
\label{eq:continuous_spatial_phenotype_separable}
\end{equation}
where $A(\theta,\theta')$ is the phenotype interaction kernel. We assume that the spatial interaction kernel satisfies $K(s)\geq0$, so that the sign of the interaction is determined entirely by the phenotype kernel. Positive values of $A_{ij}$ or $A(\theta,\theta')$ therefore correspond to attraction, whilst negative values correspond to repulsion. The fully attractive case is the most common in applications such as aggregation and cell--cell adhesion, whereas repulsive interactions arise naturally in contexts such as contact inhibition or fleeing \cite{painter2015nonlocal, jewell2025chase, potts2019_memory}.

Thus, with the above assumption, phenotype determines the strength and direction of interactions, but not their dependence on physical separation. This is natural when phenotypes differ in their overall adhesive, attractive, or signalling strength whilst sharing a characteristic interaction length scale. For example, cells may differ in adhesive strength through differences in cadherin expression whilst interacting over a common range set by cell--cell contact \cite{foty_steinberg2005differential_adhesion}. Similarly, through differences in receptor expression, chemotaxing cells can vary in their sensitivity despite responding to the same signalling molecule, and hence over the same spatial scale \cite{chang2015cell_receptor_differ,salek2019bactera_vary_sensitivity,lorenzi_painter2025phenotype_chemotaxis}. Even when different extracellular signalling molecules are involved, those of comparable size can have similar diffusivities \cite{fleury2006_similar_diffusivities}. Likewise, individual animals within a group may differ in their responsiveness to neighbours whilst using the same sensory cues to coordinate their movement \cite{herbert2016_animal_movement}. Other choices are possible; for example, Lorenzi et al.~\cite{lorenzi2024_nonlocal_kinetic_phenotype} instead assume that phenotype determines the interaction length scale.

For convenience, we introduce overall interaction strength parameters, $\tilde{\mu}>0$ in the discrete formulation and $\mu>0$ in the continuous formulation, and, without loss of generality, choose the normalisation
\[
\left(\frac{|V|}{N}\right)^2\sum_{i=1}^N\sum_{j=1}^N A_{ij}
=
\int_V\int_V A(\theta,\theta')\,\mathrm{d}\theta'\,\mathrm{d}\theta
=
\int_\Omega K(s)\,\mathrm{d}s
\equiv 1.
\]
This separates the functional form of the interactions from their overall magnitude: the phenotype kernel $A_{ij}$ (or $A(\theta,\theta')$) determines the phenotype dependence, the spatial kernel $K(s)$ the spatial dependence, and $\tilde{\mu}$ (or $\mu$) the overall interaction strength. To ensure consistency between the discrete and continuous formulations under the correspondence \eqref{eq:W_ij_W_theta_theta'_correspondence}, we define
$\tilde{\mu}={|V|}{N^{-1}}\,\mu$.

We next assume that the phenotypic effects of the two interacting individuals are themselves separable. Hence, in the discrete formulation,
\begin{equation}
    A_{ij}=B_iC_j,
\label{eq:discrete_signal_response}
\end{equation}
while in the continuous formulation,
\begin{equation}
    A(\theta,\theta')=B(\theta)C(\theta').
\label{eq:continuous_signal_response}
\end{equation}
Here, $B$ and $C$ are referred to as the \textit{response} and \textit{signal} functions, respectively. The response function determines how strongly an individual responds (or, equivalently, its susceptibility) to an interaction, while the signal function determines how strongly an individual contributes to that interaction through its signal production. One way of interpreting \eqref{eq:discrete_signal_response}--\eqref{eq:continuous_signal_response} is as a ``no preferential treatment" assumption: interactions depend only on two independent factors, how strongly one phenotype emits a signal and how strongly the other responds to that signal. That is, individuals distinguish only the total strength of an incoming signal, not the phenotype from which it originated.

The present study (Part I) focuses on this class of separable phenotype interactions, in which phenotype represents an absolute aggregation strength. In the follow-up to this study (Part II \cite{jewell2026phenotype_part2}), we will instead consider interactions that depend only on phenotypic similarity, so that $A_{ij}=A(|i-j|)$ in the discrete formulation and $A(\theta,\theta')=A(|\theta-\theta'|)$ in the continuous formulation. These interactions possess a fundamentally different mathematical structure and biological interpretation.

Interactions with a response--signal separation structure arise naturally across different biological contexts. For example, $C$ could describe how much aggregation pheromone an animal releases, while $B$ describes how strongly another individual moves towards it. \changes{Similarly, in a cell--cell adhesion setting, $C$ could represent the abundance of membrane-bound adhesive ligands presented by a cell. Meanwhile, $B$ represents the strength of the signal induced by ligand binding to membrane-bound receptors on the responding cell, depending, for example, on receptor abundance and replenishment. The product $BC$ then provides a simple mass-action-type approximation in which interaction strength increases with both ligand availability and receptor-mediated responsiveness.}

We consider monotonic response and signal functions so that phenotype has a \changes{unique} biological interpretation. If both functions increase with phenotype, then larger phenotype values correspond to individuals that both produce stronger interaction signals and respond more strongly to signals produced by others. Alternatively, one function may increase while the other decreases, representing a trade-off between signal production and response susceptibility. Both cases are considered in this work, inspired by the phenotype-structured chemotaxis model of Lorenzi and Painter \cite{lorenzi_painter2025phenotype_chemotaxis}, in which phenotype simultaneously \changes{specifies} chemoattractant secretion and chemotactic sensitivity.

\section{Pattern formation through linear instability}
\label{sec:linear_stability}

To investigate how phenotypic heterogeneity influences aggregation, we perform a linear stability analysis of the homogeneous steady states of the discrete and continuous phenotype models \eqref{eq:discrete_model} and \eqref{eq:continuum_model}. Our first aim is to determine when small spatially heterogeneous perturbations grow, leading to the spontaneous formation of spatial patterns. We then investigate the structure of these patterns in both physical and phenotype space.

Unlike standard linear stability analyses of pattern formation, however, the presence of phenotype structure introduces an additional layer of complexity. In physical space, perturbations may still be decomposed into eigenfunctions of the Laplacian (in the present setting, Fourier modes), but in phenotype space the appropriate eigenfunctions are generally not known \textit{a priori}. Instead, they depend on both the phenotype kernel and the rate of phenotypic switching. For this reason, unlike the standard approach to linear stability analysis used in most studies of pattern-forming models, which assumes a complex exponential ansatz from the outset \cite{murray2003_pattern_formation, jewell2023_dimension}, we proceed more carefully from first principles. We therefore first derive the appropriate eigenvalue problems, before specialising to particular classes of phenotype interactions and switching regimes.

As is standard in linear stability analyses of pattern formation, the eigenvalues, $\lambda(k)$, are the growth rates of perturbations with spatial wavenumber $k$. Linear instability, and hence pattern formation, occurs whenever $\text{Re}(\lambda(k))>0$ for some $k>0$.

\subsection{Discrete phenotype formulation}

Beginning with the discrete model \eqref{eq:discrete_model}, we linearise about \changes{a} homogeneous steady state \mbox{$u_i(t,\mathbf{x})=U>0\,$}\footnote{\changes{The homogeneous state is unique and determined by the initial conditions. }For positive phenotypic switching rates and no-flux boundary conditions in phenotype space, any steady state that is homogeneous in physical space must also be homogeneous in phenotype space. The zero-switching case is discussed separately in Section~\ref{sec:slow_switching}.} by taking $u_i(t, \mathbf{x})=U+\tilde{u}_i(t, \mathbf{x})$, where $|\tilde{u}_i(t, \mathbf{x})|\ll 1$ is a small perturbation that is heterogeneous in both physical space and phenotype index. Retaining only leading order terms gives
\begin{equation}
    \frac{\partial \tilde{u}_i(t,\mathbf{x})}{\partial t}
    =
    \nabla_{\mathbf{x}}^2\tilde{u}_i(t, \mathbf{x})
    -U\,p(U)\nabla_{\mathbf{x}}^2 \int_{\Omega}
    \sum_{j=1}^N
    W_{ij}(s)\,
    \tilde{u}_j(t,\mathbf{x}+\mathbf{s})
    \,\mathrm{d}^d\mathbf{s}
    +
    \tilde{\beta}\sum_{j=1}^{N}L_{ij}\tilde{u}_j(t, \mathbf{x}),
\label{eq:discrete_model_linearised}
\end{equation}
for $i=1,\ldots,N$. We then expand the perturbation using its Fourier transform in physical space,
\[
\tilde{u}_i(t,\mathbf{x})
=
\int_{\mathbb{R}^d}
e^{i\mathbf{q}\cdot\mathbf{x}}
P_i(t,\mathbf{q})
\,\mathrm{d}^d\mathbf{q}.
\]
\changes{For simplicity, throughout the linear stability analysis we work as though $\Omega=\mathbb{R}^d$, treating the wavenumber as continuous. The same dispersion relation applies on the periodic domain $\Omega=\mathbb{T}^d$, but with wavenumbers restricted to the discrete set of admissible Fourier modes. Substituting the above transform into \eqref{eq:discrete_model_linearised} and projecting onto the Fourier mode $e^{i\mathbf{k}\cdot\mathbf{x}}$ using the orthogonality of Fourier modes yields}
\begin{align}
    \frac{\partial P_i(t,\mathbf{k})}{\partial t}
    &=
    -k^2P_i(t,\mathbf{k})
    +
    k^2Up(U)
    \int_{\mathbb{R}^d}
    \sum_{j=1}^N
    W_{ij}(s)\,
    e^{i\mathbf{k}\cdot\mathbf{s}}
    P_j(t,\mathbf{k})
    \,\mathrm{d}^d\mathbf{s}
    +
    \tilde{\beta}
    \sum_{j=1}^{N}
    L_{ij}\,
    P_j(t,\mathbf{k})
    \nonumber\\
    &=
    \sum_{j=1}^N \left[ -k^2\delta_{ij}+
    k^2\tilde{\mu}\,Up(U)\,
    F(k)\,A_{ij} +\tilde{\beta}L_{ij}\right]P_j(t, \mathbf{k})
\label{eq:discrete_fourier}
\end{align}
where the second line follows from the separability assumption
\eqref{eq:discrete_spatial_phenotype_separable}, and
\[
    F(k)
    \equiv
    \int_{\mathbb{R}^d}
    K(s)\,
    e^{i\mathbf{k}\cdot\mathbf{s}}
    \,\mathrm{d}^d\mathbf{s}
\]
is the $d$-dimensional Fourier transform of the spatial interaction kernel. Since $K(s)$ is isotropic, depending only on the separation distance $s=|\mathbf{s}|$, its Fourier transform depends only on the wavenumber magnitude $k=|\mathbf{k}|$. We assume throughout that $F(k)\geq0$, as this holds for many commonly used interaction kernels, \changes{such as an exponential decay kernel and a Gaussian kernel}, and is a reasonable assumption for interactions that weaken sufficiently with distance. Further discussion of this assumption and its relaxation is given in \cite{jewell2023_dimension, jewell2025chase}.

It follows from equation \eqref{eq:discrete_fourier} that the linear growth rates of perturbations with spatial wavenumber $k$ are given by the eigenvalues, $\lambda(k)$, of the stability matrix with elements
\begin{equation}
    M_{ij}(k)
    =
    -k^2\delta_{ij}
    +
    k^2\tilde{\mu}\,Up(U)F(k)A_{ij}
    +
    \tilde{\beta}L_{ij}.
\label{eq:discrete_stability_matrix}
\end{equation}
The challenge is therefore to determine the spectrum of $M_{ij}(k)$. Its eigenvalues govern the growth of perturbations and hence the capacity for pattern formation, while the corresponding eigenvectors determine the phenotype structure of the resulting patterns. Both depend critically on the phenotype interaction matrix $A_{ij}$ and the phenotypic switching rate $\tilde{\beta}$. We now investigate how these ideas translate to the continuous phenotype formulation.

\subsection{Continuous phenotype formulation}

The corresponding analysis for the continuous phenotype model \eqref{eq:continuum_model} proceeds similarly. Linearising about a homogeneous steady state using $u(t,\mathbf{x},\theta)=U+\tilde{u}(t,\mathbf{x},\theta)$ gives
\begin{equation}
    \frac{\partial\tilde{u}(t,\mathbf{x},\theta)}{\partial t}
    =
    \nabla_{\mathbf{x}}^2\tilde{u}(t,\mathbf{x},\theta)
    -
    U\,p(U)\nabla_{\mathbf{x}}^2
    \int_{\Omega}\int_V
    W(s,\theta,\theta')
    \tilde{u}(t,\mathbf{x}+\mathbf{s},\theta')
    \,\mathrm{d}\theta'
    \,\mathrm{d}^d\mathbf{s}
    +
    \beta
    \frac{\partial^2\tilde{u}(t,\mathbf{x},\theta)}{\partial\theta^2}.
\label{eq:continuous_model_linearised}
\end{equation}
Expanding \changes{$\tilde{u}(t,\mathbf{x},\theta)=
\int_{\mathbb{R}^d}
e^{i\mathbf{q}\cdot\mathbf{x}}
P(t,\mathbf{q}, \theta)
\,\mathrm{d}^d\mathbf{q}$} with a Fourier transform in physical space, projecting onto the Fourier mode $e^{i\mathbf{k}\cdot\mathbf{x}}$, and using the separability assumption \eqref{eq:continuous_spatial_phenotype_separable} yields
\begin{equation}
    \frac{\partial P(t,k,\theta)}{\partial t}
    =
    -k^2P(t,k,\theta)
    +
    k^2\mu\,Up(U)
    F(k)
    \int_V
    A(\theta,\theta')
    P(t,k,\theta')
    \,\mathrm{d}\theta'
    +
    \beta
    \frac{\partial^2P(t,k,\theta)}{\partial\theta^2}.
\label{eq:continuous_fourier}
\end{equation}
To determine the linear growth rates, we seek separable solutions \changes{in terms of $t$ and $\theta$}  of the form
\[
P(t,k,\theta)
=
W(t, k)\,\Phi(\theta,k),
\]
where $W(t, k)$ describes the temporal growth of the perturbation, while the phenotype eigenfunction $\Phi(\theta,k)$ describes the distribution of phenotypes within the corresponding physical space Fourier mode. Separating variables yields
\[
W(t, k)
=
W_0e^{\lambda(k)t},
\]
where $\lambda(k)$ is the growth rate, \changes{which has no $\theta$ dependence}, and $\Phi(\theta,k)$ satisfies
\begin{equation}
\lambda(k)\,
\Phi(\theta,k)
=
-k^2\Phi(\theta,k)
+
\beta
\frac{\partial^2\Phi(\theta,k)}{\partial\theta^2}
+
k^2\mu\,Up(U)
F(k)
\int_V
A(\theta,\theta')
\Phi(\theta',k)
\,\mathrm{d}\theta'.
\label{eq:continuous_eigenproblem}
\end{equation}
The no-flux boundary condition \eqref{eq:phenotype_BC} also requires that
\begin{equation}
\frac{\partial\Phi(\theta,k)}{\partial\theta}
=
0,
\qquad
\theta\in\partial V.
\label{eq:continuous_eigenproblem_BC}
\end{equation}

Where the discrete formulation reduces to determining the spectrum of a finite-dimensional matrix, the continuous formulation requires the solution of the integro-differential eigenvalue problem \eqref{eq:continuous_eigenproblem}--\eqref{eq:continuous_eigenproblem_BC}. In both formulations, the eigenvalues govern the capacity for pattern formation, while the phenotype structure of the resulting patterns is determined by the corresponding eigenvectors (discrete) or eigenfunctions (continuous). Although both formulations could be unified using operator and inner-product notation, we deliberately retain their explicit discrete and continuous forms. Doing so makes their distinct mathematical structures and associated analytical tools transparent, allowing us to use whichever formulation is more natural for a given calculation. We exploit this flexibility throughout the remainder of the paper, using the separability assumption \eqref{eq:discrete_signal_response}--\eqref{eq:continuous_signal_response} and considering different phenotypic switching regimes to make analytical progress.

\section{Slow phenotypic switching regime}
\label{sec:slow_switching}

In this section, we consider the regime of negligibly slow phenotypic switching, $\tilde{\beta}\to 0$, or equivalently $\beta\to 0$. \changes{We take this limit before the continuum phenotype limit, ensuring that switching remains negligible as $N\to\infty$.} We then solve the discrete and continuous spectral problems derived in the previous section to obtain explicit conditions for pattern formation. Finally, we compare these analytical predictions with numerical simulations and discuss the biological insight that they provide.
\subsection{Pattern formation conditions}

\subsubsection{Discrete formulation}

In the limit $\tilde{\beta}\to0$, the stability matrix \eqref{eq:discrete_stability_matrix} reduces to
\begin{equation}
    M_{ij}(k)
    =
    -k^2\delta_{ij}
    +
    k^2\tilde{\mu}\,Up(U)F(k)A_{ij}.
\end{equation}
Since the first term is proportional to the identity matrix, determining the growth rates reduces to finding the eigenvalues of the phenotype interaction matrix $A_{ij}$.

Using the response--signal separation assumption \eqref{eq:discrete_signal_response},
\[
A_{ij}=B_iC_j,
\]
the phenotype interaction matrix is the outer product of two vectors and therefore has rank one \cite{strang2022_linear_algebra_book}. Furthermore,
\begin{equation}
    \sum_{j=1}^N A_{ij}B_j
    =
    \sum_{j=1}^N B_iC_jB_j
    =
    \left(\sum_{j=1}^N B_jC_j\right)B_i,
\end{equation}
for $i=1,\ldots,N$, showing that $B_i$ is an eigenvector of $A_{ij}$ with eigenvalue
\[
\lambda_A
=
\sum_{j=1}^N B_jC_j.
\]
Since $A_{ij}$ has rank one, this is its only non-zero eigenvalue, while the remaining $N-1$ eigenvalues are zero. Consequently, the dispersion relation consists of a single potentially pattern-forming branch,
\begin{equation}
    \lambda(k)
    =
    -k^2
    +
    k^2\tilde{\mu}\,Up(U)F(k)
    \sum_{j=1}^N B_jC_j,
\label{eq:slow_switching_dispersion_discrete}
\end{equation}
while the remaining $N-1$ branches satisfy
\[
\lambda(k)=-k^2<0,
\]
and therefore do not contribute to pattern formation.

\subsubsection{Continuous formulation}

We now show how the above analysis is mirrored in the continuous phenotype formulation. In the limit $\beta\to0$, the phenotype eigenvalue problem \eqref{eq:continuous_eigenproblem} reduces to
\begin{equation}
\lambda(k)\,
\Phi(\theta,k)
=
-k^2\Phi(\theta,k)
+
k^2\mu\,Up(U)
F(k)
\int_V
A(\theta,\theta')
\Phi(\theta',k)
\,\mathrm{d}\theta'.
\label{eq:continuous_slow_eigenproblem}
\end{equation}
Using the response--signal separation assumption \eqref{eq:continuous_signal_response},
\[
A(\theta,\theta')
=
B(\theta)C(\theta'),
\]
we then have
\begin{equation}
\left[\lambda(k)+k^2\right]
\Phi(\theta,k)
=
\left[k^2\mu\,Up(U)F(k)
\left(
\int_V
C(\theta')
\Phi(\theta',k)
\,\mathrm{d}\theta'
\right)
\right]B(\theta).
\label{eq:continuous_slow_separable}
\end{equation}
Since the right-hand side is proportional to the response function $B(\theta)$, any non-trivial eigenfunction satisfying $\lambda(k)\neq-k^2$ must also be proportional to $B(\theta)$. Taking $\Phi(\theta,k)=B(\theta)$ and substituting this into \eqref{eq:continuous_slow_separable} gives
\begin{equation}
\lambda(k)
=
-k^2
+
k^2\mu\,Up(U)F(k)
\int_V
B(\theta)C(\theta)
\,\mathrm{d}\theta.
\label{eq:slow_switching_dispersion_continuous}
\end{equation}
This is the continuous analogue of the discrete dispersion relation \eqref{eq:slow_switching_dispersion_discrete}, with the finite sum over phenotypes replaced by an integral over phenotype space.

\subsection{Biological insight}
\label{sec:slow_switching_insight}

\begin{figure}
    \centering
    \includegraphics[width=0.9\linewidth]{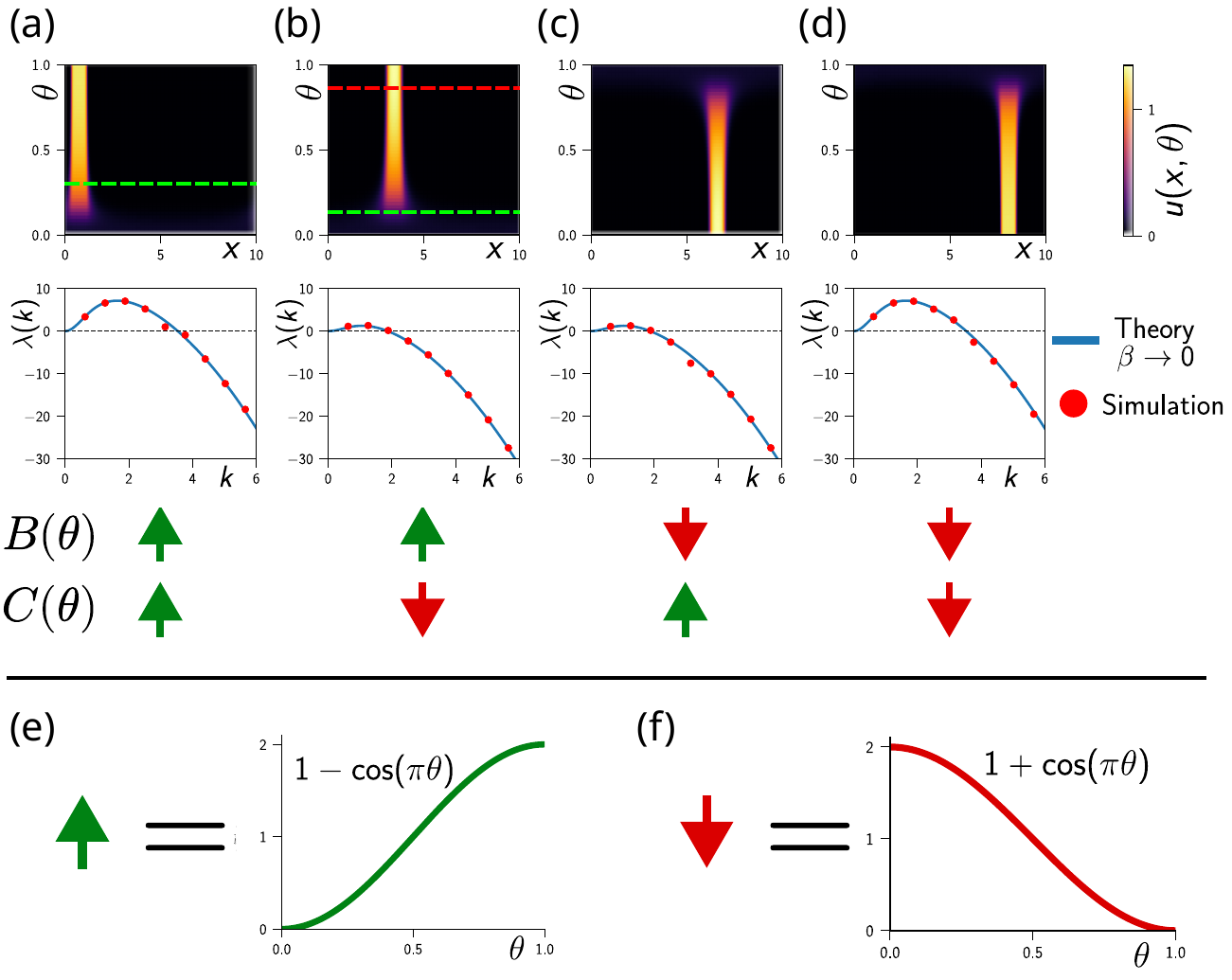}
    \caption{
    Numerical simulations in the slow phenotypic switching regime ($\beta=0.0001$). Simulations with $\beta=0$ are visually indistinguishable \changes{(not shown)}. Each of the columns (a)--(d) corresponds to a different combination of increasing or decreasing response and signal functions. For each case, the top panel shows the late-time $(t=500)$ solution $u(x,\theta)$, with physical space $x$ on the horizontal axis and phenotype $\theta$ on the vertical axis. In (a), all phenotypes above the green dashed horizontal line would form patterns in isolation. In (b), the green and red dashed horizontal lines mark the lower and upper thresholds, respectively, so that only phenotypes between the two thresholds would form patterns in isolation. Nevertheless, \changes{in all cases,} phenotypes outside these ranges are recruited into the collective pattern through coupling to the rest of the population. The lower panels compare the dispersion relation predicted by \eqref{eq:slow_switching_dispersion_discrete} \& \eqref{eq:slow_switching_dispersion_continuous} with that inferred directly from the simulations using dynamic mode decomposition. The icons below each column indicate whether the response function $B$ and signal function $C$ increase or decrease with phenotype. Panels (e) and (f) show the functional forms used for increasing and decreasing response or signal functions, namely $1-\cos(\pi\theta)$ and $1+\cos(\pi\theta)$, respectively. Parameters: $N=50$, $\mu=100$, $U=0.1$.
    }
    \label{fig:slow_switching_main}
\end{figure}

\begin{figure}
    \centering
    \includegraphics[width=0.9\linewidth]{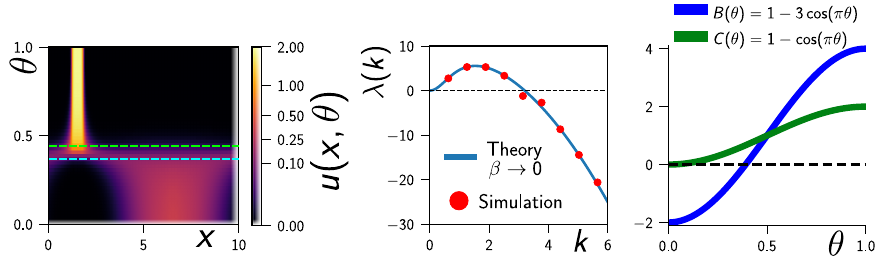}
    \caption{
    Numerical simulation illustrating \textbf{out-of-phase} aggregation when the response function changes sign. The response and signal functions are shown in the third panel, with $B(\theta)=1-3\cos(\pi\theta)$, so that phenotypes with $\theta\lesssim~0.392$ have negative response, thereby repelling from all phenotypes, including themselves. The first panel shows the late-time ($t=500$) solution $u(x,\theta)$, while the second compares the dispersion relation predicted by \eqref{eq:slow_switching_dispersion_discrete} and \eqref{eq:slow_switching_dispersion_continuous} with that inferred directly from the simulation using dynamic mode decomposition. In the first panel, phenotypes below the green dashed line would not form patterns in isolation, while those below the cyan dashed line are additionally repulsive. Nevertheless, coupling between phenotypes drives the entire population into a collective pattern, with the repulsive phenotypes aggregating out of phase with the attracting phenotypes. Parameters: $N=50$, $\mu=50$, $U=0.1$.
    }
    \label{fig:slow_switching_sin+-}
\end{figure}

\begin{figure}
    \centering
    \includegraphics[width=0.9\linewidth]{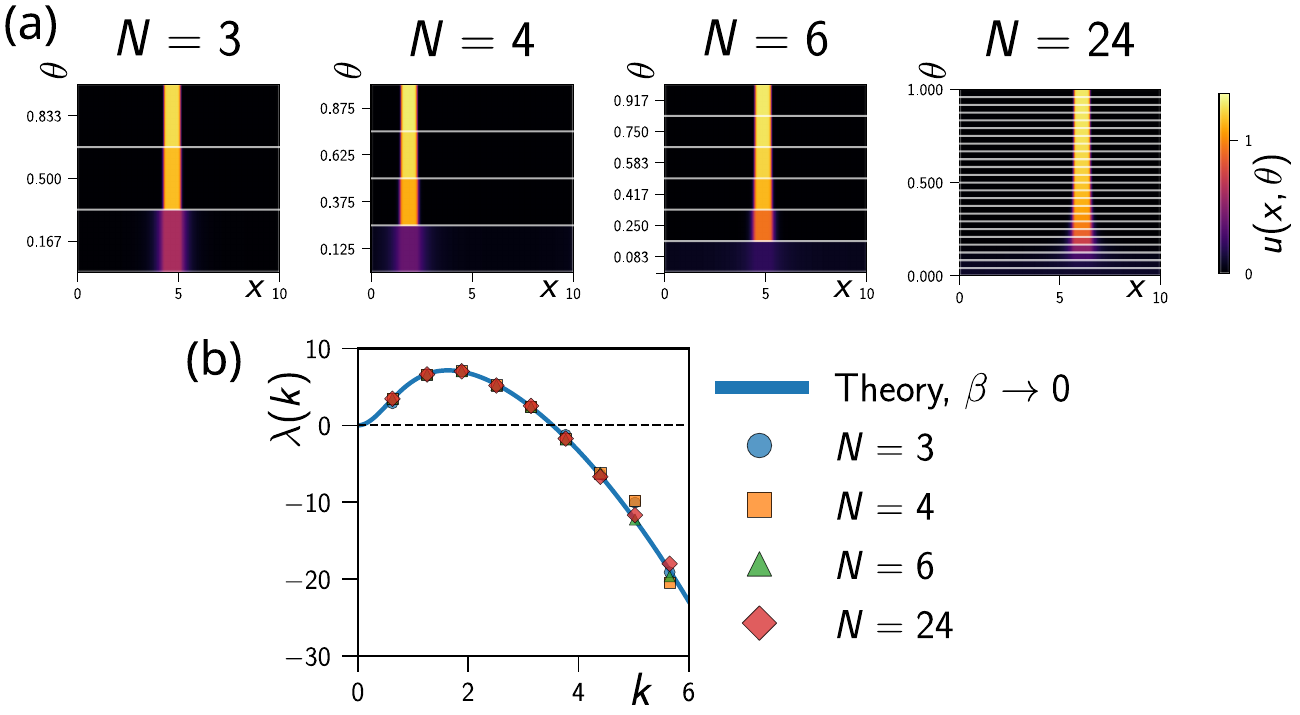}
    \caption{Results from simulations using the same parameters as Figure~\ref{fig:slow_switching_main}(a), but with the phenotype discretisation coarsened from $N=50$ to $N=3$, $N=4$, $N=6$, and $N=24$. \changes{Gray horizontal lines delimit the different phenotype classes.} Each heatmap in (a) shows the late-time $(t=500)$ solution $u(x,\theta)$. Differences in aggregate location arise only from different realisations of the random initial conditions, since the periodic system is translationally invariant and therefore has no preferred location. Panel (b) compares the dispersion relation predicted by \eqref{eq:slow_switching_dispersion_discrete} with that inferred directly from each simulation using dynamic mode decomposition. The close agreement between the theory and the simulations for every $N$ illustrates that the discrete theory remains accurate for both small and large $N$. Parameters: $\mu=100$, $U=0.1$.
    }
    \label{fig:slow_switching_small_N}
\end{figure}

We now interpret the biological insights provided by the theory in the slow phenotypic switching regime and supplement these predictions with numerical simulations.

\changes{Although the analytical calculations use both discrete and continuous phenotype formulations, all numerical simulations are performed by discretising the continuous phenotype variable, allowing comparison with both theories.} We therefore present all simulation results using the continuous notation \(u(x,\theta)\), \(\mu\), and \(\beta\), together with the number of phenotype classes \(N\), even when comparing with the discrete theory. Full details of the numerical implementation, \changes{initial conditions}, and parameters are given in Appendix~\ref{sec:appendix_numerics}.

Despite the complexity of having $N$ phenotypes, or even a continuum of phenotypes, the structure imposed by response--signal separation greatly simplifies the pattern-forming dynamics: for each spatial wavenumber $k$, there is only one phenotype mode that can drive pattern formation. Both its phenotype structure and growth rate can be characterised explicitly, with the latter given by the dispersion relations~\eqref{eq:slow_switching_dispersion_discrete} and \eqref{eq:slow_switching_dispersion_continuous}.

We see that the growth rate, $\lambda(k)$, depends on the inner product of the response and signal functions, $B$ and $C$. Pattern formation is therefore promoted when signal production and response are positively correlated, so that the individuals that produce stronger signals are also those that respond more strongly. Conversely, negative correlation, in which phenotypes specialise towards either signal production or response, suppresses pattern formation. This is illustrated in Figure~\ref{fig:slow_switching_main}, where the positively correlated cases (a) and (d) have larger growth rates than the negatively correlated cases (b) and (c). The theoretical dispersion relations closely agree with those inferred directly from the simulations using dynamic mode decomposition (DMD; Appendix~\ref{sec:appendix_dmd}).

The phenotype structure of the resulting patterns has an even simpler dependence: the phenotype eigenvector or eigenfunction is proportional to the response function $B$. Thus, $B$ determines how strongly each phenotype participates in the pattern. In contrast, how much signal each phenotype produces, $C$, still affects the growth rate but has no effect on the structure of the pattern. In Figure~\ref{fig:slow_switching_main}(a,b), where $B$ increases with phenotype, the more responsive phenotypes aggregate with greater amplitude; this structure is reversed in (c,d), where $B$ decreases. Comparing (a) with (b), or (c) with (d), shows that changing $C$ does not alter this phenotype structure.

Comparing with a model containing only a single phenotype, we see that phenotypic heterogeneity allows weakly interacting phenotypes to participate in pattern formation. Phenotypes that would be unable to pattern in isolation can nevertheless be incorporated into a collective pattern through their interactions with the rest of the population. Examples include phenotypes below the green dashed lines in Figure~\ref{fig:slow_switching_main}(a,b), whose response is too weak to support pattern formation in isolation, and those above the red dashed line in Figure~\ref{fig:slow_switching_main}(b), whose signal production is too weak. This collective patterning even extends to phenotypes that are themselves repulsive.

To illustrate this, in Figure~\ref{fig:slow_switching_sin+-} we allow the response function $B$ to change sign. Phenotypes below the critical value $\theta_c\approx0.392$ therefore respond by moving away from all phenotypes, including themselves. Since the phenotype eigenfunction is proportional to $B$, these repulsive phenotypes nevertheless aggregate as part of the collective pattern, but out of phase with the attracting phenotypes\changes{, with density peaks at different spatial locations.} In contrast, in cases where $B$ remains positive while $C$ changes sign, the phenotype eigenfunction remains positive and all phenotypes aggregate in phase.

An important additional result is that the slow-switching dispersion relations are purely real, so linear instability cannot generate temporal oscillations under the response--signal separation assumption. This does not preclude oscillations arising through nonlinear mechanisms. Examples of such behaviour are presented later in Section~\ref{sec:spatio-temporal_oscillations}, although we have only observed them at intermediate phenotypic switching rates.

None of the above conclusions rely on having a finely resolved phenotype space. The discrete theory applies equally to small $N$, as illustrated in Figure~\ref{fig:slow_switching_small_N}, where choosing 3, 4, 6, 24, or 50 phenotype classes reproduces the same qualitative behaviour and shows close agreement between the theoretical and numerically inferred dispersion relations.

For completeness, we also extend our analysis beyond steady states that are homogeneous in phenotype, $U$, to those with an arbitrary phenotype distribution, $U_i$ or $U(\theta)$. For any $\beta>0$, phenotypic switching enforces homogeneity, whereas at strictly zero switching, $\beta=0$, spatially homogeneous steady states may retain an arbitrary phenotype distribution. Repeating the linear stability analysis about such states simply rescales the response function according to $B_i\mapsto U_iB_i$ or $B(\theta)\mapsto U(\theta)B(\theta)$\changes{, with $p(U)\mapsto p(\overline{U})$}. Thus, although the zero-switching case can retain dependence on the initial conditions, this introduces no fundamentally new linear behaviour and can be readily accounted for within the existing analysis.

\section{Fast phenotypic switching regime}
\label{sec:fast_switching}

In this section, we consider the regime of asymptotically fast phenotypic switching, $\tilde{\beta}\to\infty$, or equivalently $\beta\to\infty$. In this regime, phenotypic switching occurs on a much faster timescale than movement in physical space, rapidly homogenising the phenotype distribution. We determine the corresponding pattern formation conditions for the discrete and continuous phenotype formulations, before comparing the analytical predictions with numerical simulations and discussing the biological insight that they provide.

\subsection{Pattern formation conditions}

\subsubsection{Discrete formulation}

Unlike the slow switching limit, the fast switching limit cannot be obtained simply by taking the limit of the stability matrix. Instead, we consider the limit $\tilde{\beta}\to\infty$ and expand the Fourier amplitudes in powers of the inverse switching rate,
\[
P_i(t,k)
=
P_i^{(0)}(t,k)
+
\tilde{\beta}^{-1}P_i^{(1)}(t,k)
+
\mathcal{O}(\tilde{\beta}^{-2}).
\]
At leading order, the phenotypic switching term in \eqref{eq:discrete_fourier} requires
\begin{equation}
    \sum_{j=1}^N L_{ij}P_j^{(0)}(t,k)=0,
\end{equation}
for $i=1,\ldots,N$. Thus, the leading-order Fourier amplitudes lie in the nullspace of the graph Laplacian. From \eqref{eq:graph_laplacian}, this implies
\[
P_1^{(0)}=P_2^{(0)},\qquad
P_{N-1}^{(0)}=P_N^{(0)},\qquad
P_{i+1}^{(0)}-P_i^{(0)}
=
P_i^{(0)}-P_{i-1}^{(0)},
\qquad
i=2,\ldots,N-1.
\]
The only solution is the constant vector
\[
P_1^{(0)}=P_2^{(0)}=\cdots=P_N^{(0)}
\equiv P_0(t,k).
\]
Physically, this corresponds to rapid phenotypic switching homogenising each spatial Fourier mode across phenotype space.

At the next order, substituting the expansion into \eqref{eq:discrete_fourier} gives
\begin{equation}
\frac{\partial P_0(t,k)}{\partial t}
=
-k^2P_0(t,k)
+
k^2\tilde{\mu}\,Up(U)F(k)
\sum_{j=1}^N A_{ij}P_0(t,k)
+
\sum_{j=1}^N L_{ij}P_j^{(1)}(t,k).
\label{eq:fast_switching_discrete_next_order}
\end{equation}
Summing \eqref{eq:fast_switching_discrete_next_order} over all phenotypes and using the fact that the column sums of the graph Laplacian vanish eliminates the unknown correction $P_j^{(1)}$, giving
\begin{equation}
    \frac{\partial P_0(t,k)}{\partial t}
    =
    -k^2P_0(t,k)
    +
    k^2\tilde{\mu}\,Up(U)F(k)
    \frac{1}{N}
    \sum_{i=1}^N
    \sum_{j=1}^N
    A_{ij}
    P_0(t,k).
\label{eq:fast_switching_discrete_final_ODE}
\end{equation}
This summation can equivalently be understood as imposing the solvability condition for the next-order equation. Since the graph Laplacian is symmetric and has the constant vector as its zero eigenvector, the Fredholm alternative requires the terms not involving $P_j^{(1)}$ in \eqref{eq:fast_switching_discrete_next_order} to be orthogonal to this vector, i.e.\ they must sum to zero.

Equation\,\eqref{eq:fast_switching_discrete_final_ODE} implies that the leading-order fast-switching dispersion relation is
\begin{equation}
    \lambda(k)
    =
    -k^2
    +
    k^2\tilde{\mu}\,Up(U)F(k)
    \frac{1}{N}
    \sum_{i=1}^N
    \sum_{j=1}^N
    A_{ij}.
\label{eq:fast_switching_dispersion_discrete}
\end{equation}
With the response--signal separation assumption \eqref{eq:discrete_signal_response},
\[
A_{ij}=B_iC_j,
\]
this simplifies to
\begin{equation}
    \lambda(k)
    =
    -k^2
    +
    k^2\tilde{\mu}\,Up(U)F(k)
    \frac{1}{N}
    \left(
    \sum_{i=1}^N B_i
    \right)
    \left(
    \sum_{j=1}^N C_j
    \right).
\label{eq:fast_switching_dispersion_discrete_separable}
\end{equation}

\subsubsection{Continuous formulation}

We now show how the above analysis is mirrored in the continuous phenotype formulation. As $\beta\to\infty$, we similarly expand the phenotype eigenfunction
\[
\Phi(\theta,k)
=
\Phi^{(0)}(\theta,k)
+
\beta^{-1}\Phi^{(1)}(\theta,k)
+
\mathcal{O}(\beta^{-2}).
\]
At leading order, the phenotypic switching term in \eqref{eq:continuous_eigenproblem} requires
\begin{equation}
    \frac{\partial^2\Phi^{(0)}(\theta,k)}{\partial\theta^2}=0,
\end{equation}
subject to the no-flux boundary conditions \eqref{eq:continuous_eigenproblem_BC}. The only solution is the $\theta$-independent function
\begin{equation}
    \Phi^{(0)}(\theta,k)\equiv\Phi_0(k).
\end{equation}
As in the discrete formulation, rapid phenotypic switching homogenises the phenotype eigenfunction.

At the next order, integrating \eqref{eq:continuous_eigenproblem} over phenotype space and using the no-flux boundary conditions eliminates the phenotype diffusion term containing the unknown correction $\Phi^{(1)}$, giving
\begin{equation}
    \lambda(k)
    =
    -k^2
    +
    k^2\mu\,Up(U)F(k)
    \frac{1}{|V|}
    \int_V\int_V
    A(\theta,\theta')
    \,\mathrm{d}\theta'
    \,\mathrm{d}\theta.
\label{eq:fast_switching_dispersion_continuous}
\end{equation}
Again, this step can equivalently be understood as imposing the Fredholm solvability condition.

With the response--signal separation assumption \eqref{eq:continuous_signal_response},
\[
A(\theta,\theta')
=
B(\theta)C(\theta'),
\]
the dispersion relation simplifies to
\begin{equation}
    \lambda(k)
    =
    -k^2
    +
    k^2\mu\,Up(U)F(k)
    \frac{1}{|V|}
    \left(
    \int_VB(\theta)\,\mathrm{d}\theta
    \right)
    \left(
    \int_VC(\theta)\,\mathrm{d}\theta
    \right).
\label{eq:fast_switching_dispersion_continuous_separable}
\end{equation}
This is the continuous analogue of the discrete dispersion relation \eqref{eq:fast_switching_dispersion_discrete_separable}, with finite sums over phenotype replaced by integrals over phenotype space.

\begin{figure}
    \centering
    \includegraphics[width=0.9\linewidth]{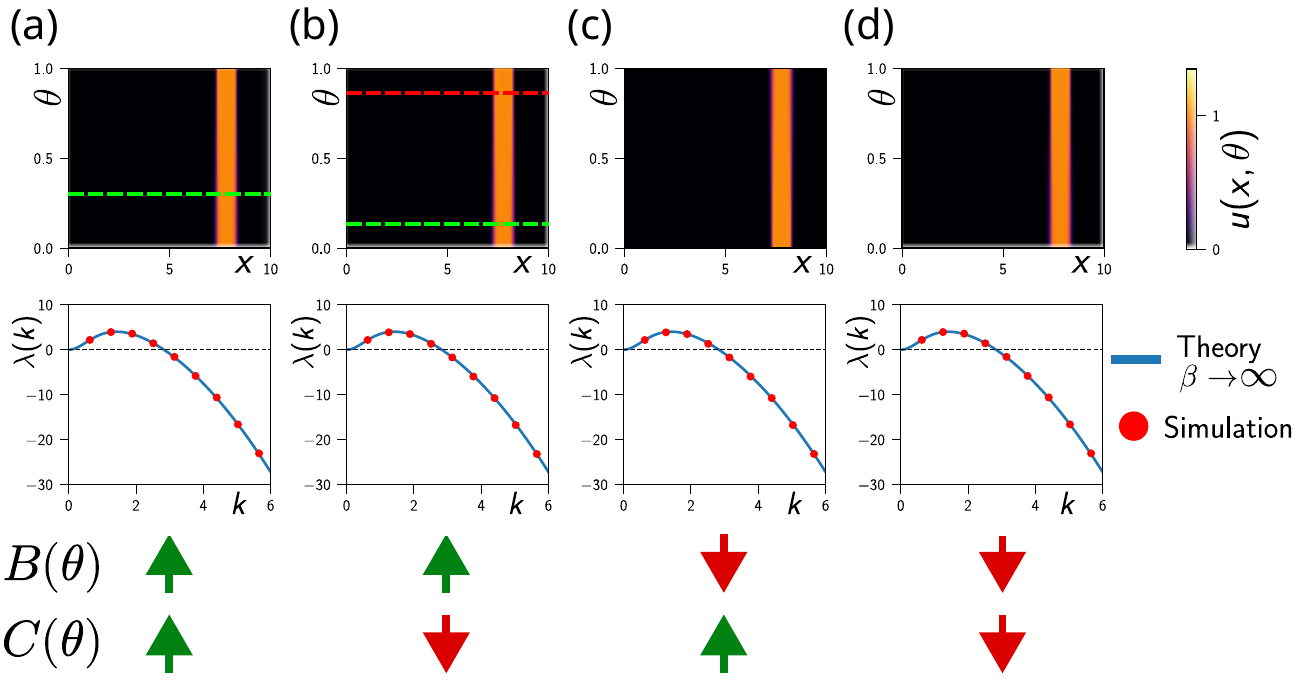}
    \caption{
    Numerical simulations in the fast phenotypic switching regime ($\beta=50$). Each of the columns (a)--(d) corresponds to a different combination of increasing ($1-\cos(\pi\theta)$) or decreasing ($1+\cos(\pi\theta)$) response and signal functions, indicated by the icons below each column. For each case, the top panel shows the late-time ($t=500$) solution $u(x,\theta)$, with physical space $x$ on the horizontal axis and phenotype $\theta$ on the vertical axis. In (a), all phenotypes above the green dashed horizontal line would form patterns in isolation. In (b), the green and red dashed horizontal lines mark the lower and upper thresholds, respectively, so that only phenotypes between the two thresholds would form patterns in isolation. In contrast to Figure~\ref{fig:slow_switching_main}, the patterns are homogeneous in phenotype, reflecting the rapid mixing induced by fast phenotypic switching. The lower panels compare the dispersion relation predicted by \eqref{eq:fast_switching_dispersion_discrete_separable} and \eqref{eq:fast_switching_dispersion_continuous_separable} with that inferred directly from the simulations using dynamic mode decomposition. As predicted by the theory, the dispersion relation is identical in all four cases. Parameters: $N=50$, $\mu=100$, $U=0.1$.
    }
    \label{fig:fast_switching_main}
\end{figure}

\begin{figure}
    \centering
    \includegraphics[width=0.9\linewidth]{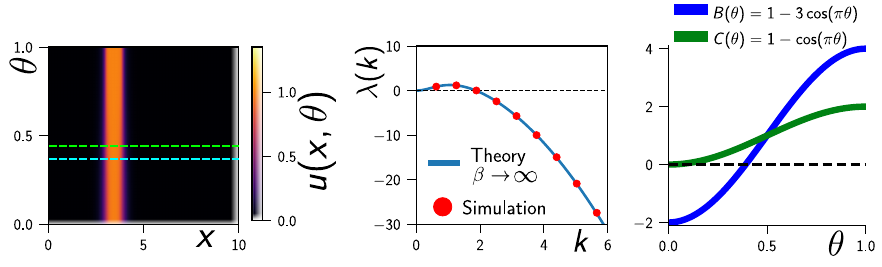}
    \caption{
    Numerical simulation illustrating \textbf{in-phase} aggregation under fast phenotypic switching ($\beta=50$) when the response function changes sign. The response and signal functions are shown in the third panel, with $B(\theta)=1-3\cos(\pi\theta)$, so that phenotypes with $\theta\lesssim0.392$ have negative response, thereby repelling from all phenotypes, including themselves. The first panel shows the late-time ($t=500$) solution $u(x,\theta)$, while the second compares the dispersion relation predicted by \eqref{eq:fast_switching_dispersion_discrete_separable} and \eqref{eq:fast_switching_dispersion_continuous_separable} with that inferred directly from the simulation using dynamic mode decomposition. In the first panel, phenotypes below the green dashed line would not form patterns in isolation, while those below the cyan dashed line are additionally repulsive. Nevertheless, rapid phenotypic switching drives the entire population into a pattern that is uniform across phenotype space, with even the repulsive phenotypes aggregating in phase with the attracting phenotypes. Parameters: $N=50$, $\mu=50$, $U=0.1$.
    }
    \label{fig:fast_switching_sin+-}
\end{figure}

\subsection{Biological insight}

We now interpret the biological insights provided by the theory in the fast phenotypic switching regime and compare its predictions with numerical simulations. A central consequence of rapid phenotypic switching is that the pattern-forming dynamics again reduce to a single phenotype mode. In contrast to the slow-switching regime, however, this simplification does not arise from or require response--signal separation. Instead, rapid switching homogenises the population across phenotype space, irrespective of the structure of the phenotype interactions.

The dispersion relations~\eqref{eq:fast_switching_dispersion_discrete} and \eqref{eq:fast_switching_dispersion_continuous} show that the growth rate, $\lambda(k)$, and hence the capacity for pattern formation, is determined only by the total interaction strength in the system, rather than by how those interactions are structured between phenotypes. Here, we use total interaction strength to refer to the double sum of $A_{ij}$, or equivalently the double integral of $A(\theta,\theta')$. Under response--signal separation, equations~\eqref{eq:fast_switching_dispersion_discrete_separable} and \eqref{eq:fast_switching_dispersion_continuous_separable} show that the growth rate therefore depends only on the total response and total signal, not on their correlation. This is illustrated in Figure~\ref{fig:fast_switching_main}: despite the different combinations of increasing and decreasing $B$ and $C$, all four cases have the same dispersion relation, with close agreement between theory and that inferred from simulations using DMD. This contrasts with the slow-switching regime, where correlations between response and signal strongly influence pattern formation.

Rapid switching homogenises the resulting patterns across phenotype space. Since the leading-order phenotype eigenvector or eigenfunction is constant, all phenotypes aggregate with equal amplitude, as seen in Figure~\ref{fig:fast_switching_main}. Consequently, phenotypes that would be unable to pattern in isolation are fully incorporated into the collective pattern, in contrast to the slow-switching regime, where less responsive phenotypes aggregate with smaller amplitude. This extends even to repulsive phenotypes: in Figure~\ref{fig:fast_switching_sin+-}, rapid switching causes the repulsive and attracting phenotypes to aggregate in phase, in contrast to the out-of-phase aggregation observed under slow switching.

As in the slow-switching regime, the dispersion relation is purely real, and so linear instability cannot generate temporal oscillations. We have also verified numerically that the fast-switching theory remains accurate for small $N$ \changes{(results not shown)}, consistent with the discrete formulation.

\section{Generic phenotypic switching}
\label{sec:generic_switching}

Having analysed the limiting cases of slow and fast phenotypic switching, we now tackle the more generic case where $0<\tilde{\beta}<\infty$, or equivalently $0<\beta<\infty$. In this regime, calculating the phenotype eigenfunctions and the corresponding growth rates is substantially more involved than in the previous limiting cases. Indeed, analysis of phenotype-structured models with generic switching rates has been identified as particularly challenging \cite{lorenzi_painter2025phenotype_chemotaxis}, with some studies instead focusing on limiting cases \cite{cusseddu_lorenzi_2026phenotype_SKT,lorenzi2024_nonlocal_kinetic_phenotype}. To overcome these analytical challenges, having the choice between the discrete and continuous phenotype formulations becomes particularly valuable, allowing whichever mathematical tools are more familiar or convenient to be used. We illustrate this with two calculations, each of which naturally lends itself to a different formulation. Together, these demonstrate that substantial analytical progress is still possible \changes{in some cases}, providing further insight into how phenotypic switching influences both the capacity for pattern formation and whether the resulting patterns can be oscillatory.

\subsection{Pattern formation}
\label{sec:sub_generic_switching_PF}

\subsubsection{General dispersion relation}

To derive the dispersion relation governing pattern formation in the generic switching regime, we use the continuous phenotype formulation.
With the response--signal separation assumption \eqref{eq:continuous_signal_response},
\[
A(\theta,\theta')=B(\theta)C(\theta'),
\]
the phenotype eigenvalue problem \eqref{eq:continuous_eigenproblem} can be written as
\begin{equation}
\left[(k^2+\lambda(k))-\beta\frac{\partial^2}{\partial\theta^2}\right]\Phi(\theta,k)
=
k^2\mu\,Up(U)F(k)\,S(k)\,B(\theta),
\label{eq:generic_switching_continuous_ode}
\end{equation}
where we have defined
\[
S(k)=\int_V C(\theta')\Phi(\theta',k)\,\mathrm{d}\theta'.
\]
\changes{Fixing $k$ and} treating $S(k)$ as a constant, equation \eqref{eq:generic_switching_continuous_ode} is a linear second-order ODE in $\theta$ with constant coefficients and forcing proportional to $B(\theta)$. This ODE can be solved directly for particular choices of $B$, with its solution then used to solve for $\lambda(k)$ for particular choices of $C(\theta)$. Instead, we first take a more general Green's function approach, and only later specify $B$ and $C$. This leads to an implicit condition for the growth rates through the self-consistency relation defining $S(k)$.

Let $G_\lambda(\theta,\theta')$ denote the Green's function for the differential operator
\[
\mathcal{L}_\lambda
=
(k^2+\lambda(k))
-
\beta\frac{\partial^2}{\partial\theta^2},
\]
defined by
\begin{equation}
\mathcal{L}_\lambda G_\lambda(\theta,\theta')
=
\delta(\theta-\theta'),
\qquad
\left.
\frac{\partial G_\lambda(\theta,\theta')}{\partial\theta}
\right|_{\theta\in\partial V}
=
0.
\label{eq:greens_function_definition}
\end{equation}
The solution to \eqref{eq:generic_switching_continuous_ode} is therefore
\begin{equation}
\Phi(\theta,k)
=
k^2\mu\,Up(U)F(k)\,S(k)
\int_V
G_\lambda(\theta,\theta')
B(\theta')
\,\mathrm{d}\theta'.
\label{eq:generic_switching_green_solution}
\end{equation}
Already, this solution suggests that the phenotype eigenfunction is a smoothed version of the response function, $B$, with the extent of the smoothing depending on, among other things, the switching rate, $\beta$. This provides a natural intermediary between the slow switching regime, where the phenotype eigenfunction is exactly $B$, and the fast switching regime, where the phenotype eigenfunction is averaged to a constant.

Substituting \eqref{eq:generic_switching_green_solution} into the definition of $S(k)$ gives
\begin{equation}
\begin{split}
S(k)
&=
k^2\mu\,Up(U)F(k)\,S(k)
\int_V
C(\theta)
\left(
\int_V
G_\lambda(\theta,\theta')
B(\theta')
\,\mathrm{d}\theta'
\right)
\mathrm{d}\theta.
\end{split}
\end{equation}
Hence, non-trivial solutions satisfy
\begin{equation}
1
=
k^2\mu\,Up(U)F(k)\,\mathcal{I}_\lambda[B,C]
\,,
\label{eq:generic_switching_continuous_dispersion}
\end{equation}
where we have defined the integral
\begin{equation}
    \mathcal{I}_\lambda[B,C]=\int_V
\int_V
C(\theta)
G_\lambda(\theta,\theta')
B(\theta')
\,\mathrm{d}\theta'
\,\mathrm{d}\theta.
\label{eq:I_lambda}
\end{equation}
To determine the growth rates, we derive the Green's function and substitute it into \eqref{eq:generic_switching_continuous_dispersion}--\eqref{eq:I_lambda}. We then specialise to particular choices of the signal and response functions, $B$ and $C$, compute the resulting integrals \eqref{eq:I_lambda}, and solve for $\lambda(k)$.

In Appendix~\ref{sec:appendix_greens_function} we derive the Green's function, showing that it takes the form
\begin{equation}
\begin{split}
G_\lambda(\theta,\theta')
=
\frac{
\cosh\!\left(
\sqrt{\frac{k^2+\lambda(k)}{\beta}}\,
\min\{\theta,\theta'\}
\right)
\cosh\!\left(
\sqrt{\frac{k^2+\lambda(k)}{\beta}}\,
\bigl(|V|-\max\{\theta,\theta'\}\bigr)
\right)
}{
\sqrt{\beta\,(k^2+\lambda(k))\,}\,
\sinh\!\left(
\sqrt{\frac{k^2+\lambda(k)}{\beta}}\,
|V|
\right)
},
\end{split}
\label{eq:greens_function}
\end{equation}
\changes{provided $\lambda(k)\neq-k^2-\beta(n\pi/|V|)^2$, where $n\in\mathbb{Z}$.}

\subsubsection{Positively correlated response and signal functions}

We first consider a case in which individuals with higher phenotype values both emit stronger signals and respond more strongly to signals. We choose the positively correlated response and signal functions
\[
B(\theta)=C(\theta)=1-\cos(\pi\theta), \qquad \theta\in V=[0,1].
\]
With these choices, the integral \eqref{eq:I_lambda} evaluates to
\[
\mathcal{I}_\lambda[B,C]
=
\frac{1}{\bigl(k^2+\lambda(k)\bigr)}
+
\frac{1}{2\bigl(k^2+\lambda(k)+\beta\pi^2\bigr)}.
\]
The calculation of this integral is given in Appendix~\ref{sec:appendix_integral_calculation}\ref{sec:appendix_integral_positive_correlation}. Substituting this into \eqref{eq:generic_switching_continuous_dispersion} and rearranging for $\lambda(k)$ yields the dispersion relation
\begin{equation}
\begin{split}
\lambda(k)
={}&
-k^2
+\frac{3}{4}k^2\mu\,Up(U)F(k)
-\frac{\beta\pi^2}{2}\\
&+
\frac{1}{4}
\sqrt{
9k^4\mu^2U^2p(U)^2F(k)^2
+
4\beta\pi^2k^2\mu\,Up(U)F(k)
+
4\beta^2\pi^4
}.
\end{split}
\label{eq:positive_correlation_dispersion}
\end{equation}
Despite the analytical challenges associated with generic switching rates, we have obtained a closed-form expression for the growth rate $\lambda(k)$. There are two branches of the dispersion relation, with the second corresponding to taking the negative square root in \eqref{eq:positive_correlation_dispersion}. However, this branch satisfies $\lambda(k)<0$ for all $k>0$ and therefore cannot drive pattern formation. We therefore focus on the positive branch.

Equation~\eqref{eq:positive_correlation_dispersion} recovers the limiting regimes derived previously. As $\beta\to0$,
\[
\lambda(k)
\to
-k^2
+\frac{3}{2}k^2\mu\,Up(U)F(k)
=
-k^2
+
k^2\mu\,Up(U)F(k)
\int_V
B(\theta)C(\theta)\,\mathrm{d}\theta,
\]
recovering \eqref{eq:slow_switching_dispersion_continuous}, whilst as $\beta\to\infty$,
\[
\lambda(k)
\to
-k^2
+k^2\mu\,Up(U)F(k)
=
-k^2
+
k^2\mu\,Up(U)F(k)
\frac{1}{|V|}
\left(
\int_VB(\theta)\,\mathrm{d}\theta
\right)
\left(
\int_VC(\theta)\,\mathrm{d}\theta
\right),
\]
recovering \eqref{eq:fast_switching_dispersion_continuous}. An important observation is that $\lambda(k)$ decreases monotonically with $\beta$ and continuously interpolates between these two limits.

\subsubsection{Negatively correlated response and signal functions}
We now consider a case in which there is a trade-off between signal production and response, such that individuals with higher phenotype respond more strongly to signals but emit less of them. We choose
\[
B(\theta)=1-\cos(\pi\theta), \qquad C(\theta)=1+\cos(\pi\theta), \qquad \theta\in V=[0,1].
\]
Here, the integral \eqref{eq:I_lambda}, whose calculation is described in Appendix~\ref{sec:appendix_integral_calculation}\ref{sec:appendix_integral_negative_correlation}, evaluates to
\[
\mathcal{I}_\lambda[B,C]
=
\frac{1}{\bigl(k^2+\lambda(k)\bigr)}
-
\frac{1}{2\bigl(k^2+\lambda(k)+\beta\pi^2\bigr)}.
\]
Substituting this into \eqref{eq:generic_switching_continuous_dispersion} and rearranging for $\lambda(k)$ yields the dispersion relation
\begin{equation}
\begin{split}
\lambda(k)
={}&
-k^2
+\frac{1}{4}k^2\mu\,Up(U)F(k)
-\frac{\beta\pi^2}{2}\\
&+
\frac{1}{4}
\sqrt{
k^4\mu^2U^2p(U)^2F(k)^2
+
12\beta\pi^2k^2\mu\,Up(U)F(k)
+
4\beta^2\pi^4
}.
\end{split}
\label{eq:negative_correlation_dispersion}
\end{equation}
We once again obtain a closed-form expression for the growth rate $\lambda(k)$. There are again two branches of the dispersion relation, with the branch corresponding to the negative square root unable to drive pattern formation.

As in the positively correlated case, taking appropriate limits of \eqref{eq:negative_correlation_dispersion} recovers the slow- and fast-switching dispersion relations. Here, however, $\lambda(k)$ \textit{increases} monotonically with $\beta$, again continuously interpolating between the two limits.

\subsubsection{Discrete equivalent}
Having used the continuous formulation to derive the dispersion relation governing pattern formation for generic phenotypic switching, we now briefly compare this calculation with its counterpart in the discrete formulation.

An equivalent equation to \eqref{eq:generic_switching_continuous_dispersion} can be found in the discrete formulation by treating $B_iC_j$ as a rank-one perturbation in \eqref{eq:discrete_stability_matrix}, giving
\begin{equation}
    1
    =
    k^2\tilde{\mu}\,Up(U)F(k)
    \sum_{i=1}^N\sum_{j=1}^N
    C_i
    \left(H^{-1}\right)_{ij}
    B_j,
\label{eq:finite_switching_discrete_dispersion}
\end{equation}
\changes{where $H$ is the $N\times N$ tridiagonal matrix with entries $H_{ij}=(k^2+\lambda(k))\delta_{ij}-\tilde{\beta}L_{ij}$.
In principle, \eqref{eq:finite_switching_discrete_dispersion} can be solved for $\lambda(k)$ by first inverting $H$.} While this is certainly possible \cite{usmani1994_tridiagonal_matrix_inversion}, it relies on tools that are perhaps less familiar than those used in the continuous analogue, where the corresponding problem reduced to solving a second-order linear ODE with constant coefficients. The Green's function approach adopted above provides a general solution, while for particular choices of $B$ and $C$, an even simpler approach based on combining the homogeneous solution with an appropriate particular integral would suffice.

\subsubsection{Biological insight}

\begin{figure}
    \centering
    \includegraphics[width=0.9\linewidth]{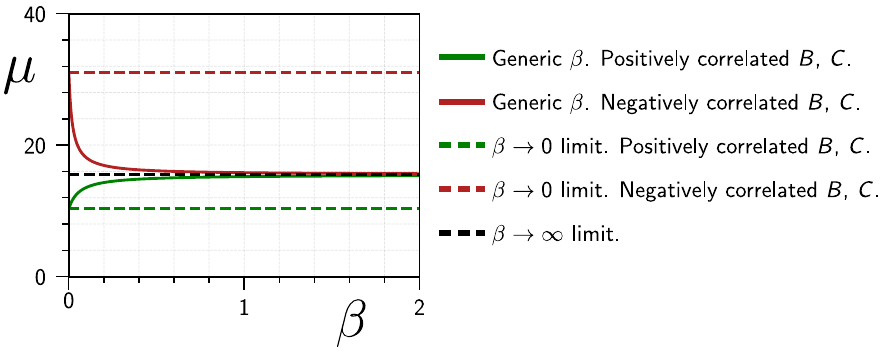}
    \caption{
    Phase diagram showing the onset of pattern formation in the $(\beta,\mu)$ plane. The bifurcation curves, defined by $\max_{k>0}\operatorname{Re}(\lambda(k))=0$, are computed from the slow-switching dispersion relation \eqref{eq:slow_switching_dispersion_continuous}, the fast-switching dispersion relation \eqref{eq:fast_switching_dispersion_continuous}, and the generic-switching dispersion relations \eqref{eq:positive_correlation_dispersion} and \eqref{eq:negative_correlation_dispersion}. Pattern formation occurs above each curve. Positive correlation between the response and signal functions lowers the critical interaction strength required for pattern formation, whereas negative correlation raises it. The generic-switching bifurcation curves interpolate continuously between the slow- and fast-switching limits, with the transition occurring over only a narrow range of switching rates.
    }
    \label{fig:bifurcation}
\end{figure}

\begin{figure}
    \centering
    \includegraphics[width=0.87\linewidth]{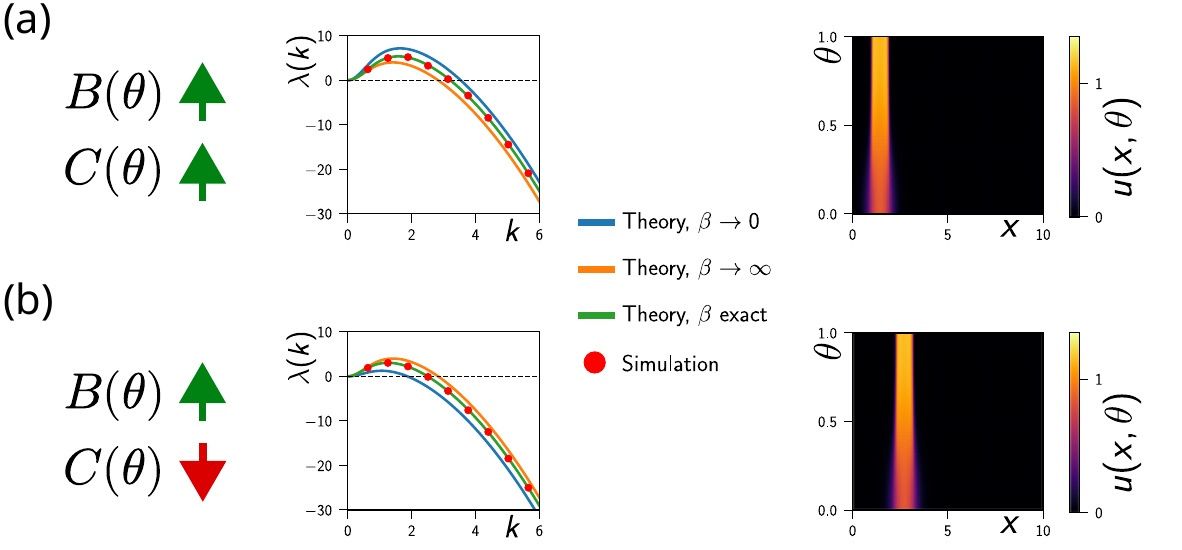}
    \caption{
    Numerical simulations at an intermediate phenotypic switching rate ($\beta=1$). Rows (a) and (b) correspond to positively correlated response and signal functions, $B(\theta)=C(\theta)=1-\cos(\pi\theta)$, and negatively correlated functions, $B(\theta)=1-\cos(\pi\theta)$ and $C(\theta)=1+\cos(\pi\theta)$, respectively, as indicated by the icons on the left. The middle panels compare the slow- and fast-switching limits with the generic-switching dispersion relations \eqref{eq:positive_correlation_dispersion} and \eqref{eq:negative_correlation_dispersion}, together with growth rates inferred directly from simulations using DMD. In both cases, the generic-switching prediction lies between the two limiting regimes and closely agrees with the simulation. Increasing the switching rate decreases the growth rate for positive correlation, while the opposite occurs for negative correlation. The right panels show the corresponding late-time ($t=500$) solutions $u(x,\theta)$. Parameters: $N=50$, $\mu=100$, $U=0.1$.
    }
    \label{fig:intermediate_switching_simulations}
\end{figure}

The above results provide a clear picture of how phenotypic switching modifies the effect of correlations between response and signal production. As seen in the slow-switching limit, positive correlation promotes pattern formation, whereas negative correlation inhibits it. Equations~\eqref{eq:positive_correlation_dispersion} and \eqref{eq:negative_correlation_dispersion} show that increasing the switching rate progressively averages out the effect of these correlations, and therefore has opposite effects in the two cases: it inhibits pattern formation when response and signal are positively correlated, but promotes it when they are negatively correlated. In the fast-switching limit, the effect of correlation is lost entirely and the two cases become equivalent.

Figure~\ref{fig:bifurcation} summarises this behaviour in the $(\beta,\mu)$ plane. The slow- and fast-switching limits bound the behaviour, while the generic-switching dispersion relations show that the growth rate varies monotonically between them as $\beta$ increases. Together, these results provide a complete classification, for the interaction structures considered here, of how phenotypic switching affects the capacity for pattern formation.

The theoretical predictions are corroborated in Figure~\ref{fig:intermediate_switching_simulations}, where the growth rates predicted by \eqref{eq:positive_correlation_dispersion} and \eqref{eq:negative_correlation_dispersion} closely match those inferred directly from the simulations. For positive correlation, the intermediate-switching growth rates lie below the slow-switching limit and above the fast-switching limit, with the ordering reversed for negative correlation.

Finally, for both choices of response and signal functions considered above, the growth rates remain purely real for all switching rates. Thus, in these cases, linear instability cannot generate spatio-temporal oscillations. Next, we ask more generally: when can spatio-temporal oscillations arise? In contrast to the above analysis, this question is more naturally addressed using the discrete phenotype formulation, \changes{as we will show below.}

\subsection{Spatio-temporal oscillations}
\label{sec:spatio-temporal_oscillations}
\subsubsection{Conditions for linear spatio-temporal oscillations}
We first show that linear spatio-temporal oscillations cannot arise when all \changes{cross-phenotype} interactions are attractive. Assuming
\[
A_{ij}>0,\quad\changes{\forall\,i\neq j}
\]
the growth rate with largest real part is real for every $k$. This result is considerably more general than those above, since it does not require the response--signal separation assumption \eqref{eq:discrete_signal_response}. The proof is simpler in the discrete formulation, where it follows directly from the Perron--Frobenius theorem. This states that any square matrix with strictly positive entries possesses a unique eigenvalue with largest real part, which is real, simple, and has an associated eigenvector whose components all have the same sign \cite{meyer2023perron_frobenius}.

For $i\neq j$, the off-diagonal entries of the stability matrix \eqref{eq:discrete_stability_matrix} satisfy
\[
M_{ij}(k)
=
k^2\tilde{\mu}\,Up(U)F(k)A_{ij}
+
\tilde{\beta}L_{ij}
>0.
\]
For the diagonal elements, we can choose a sufficiently large constant $c>0$ such that
\[
M_{ii}(k)+c>0, \qquad \forall\, i \in \{1, ..., N\}.
\]
Thus, the shifted matrix with entries
\[
M_{ij}(k)+c\delta_{ij}
\]
is strictly positive. The Perron--Frobenius theorem therefore applies, so its eigenvalue with largest real part is real, simple, and has an associated eigenvector whose components all have the same sign. Since adding $c\delta_{ij}$ merely shifts every eigenvalue by $+c$ without changing the eigenvectors, the same conclusions hold for the stability matrix \eqref{eq:discrete_stability_matrix}. Hence, the fastest-growing mode is always non-oscillatory, and every phenotype aggregates in phase.

Moreover, the only assumption required is that all \changes{cross-phenotype} interactions are attractive, as would naturally arise in applications to cell--cell adhesion. We conjecture that an analogous result also holds in the continuous formulation, although establishing it would likely require more abstract functional analysis, such as the Krein--Rutman theorem \cite{du2006_krein_rutman}. This therefore provides an example where the discrete formulation requires less heavy machinery and is arguably the more natural choice.

These results place strong restrictions on spatio-temporal oscillations arising through linear instability. Even without response--signal separation, an oscillatory instability requires some repulsive interactions, together with sufficient attraction to drive pattern formation. Under response--signal separation, our slow- and fast-switching results impose the additional requirement of an intermediate phenotypic switching rate. These are necessary but not sufficient conditions \changes{for the presence of an oscillation}. Moreover, these conclusions concern only linear mechanisms and do not preclude oscillations generated by nonlinear dynamics.

\begin{figure}
    \centering
    \includegraphics[width=0.99\linewidth]{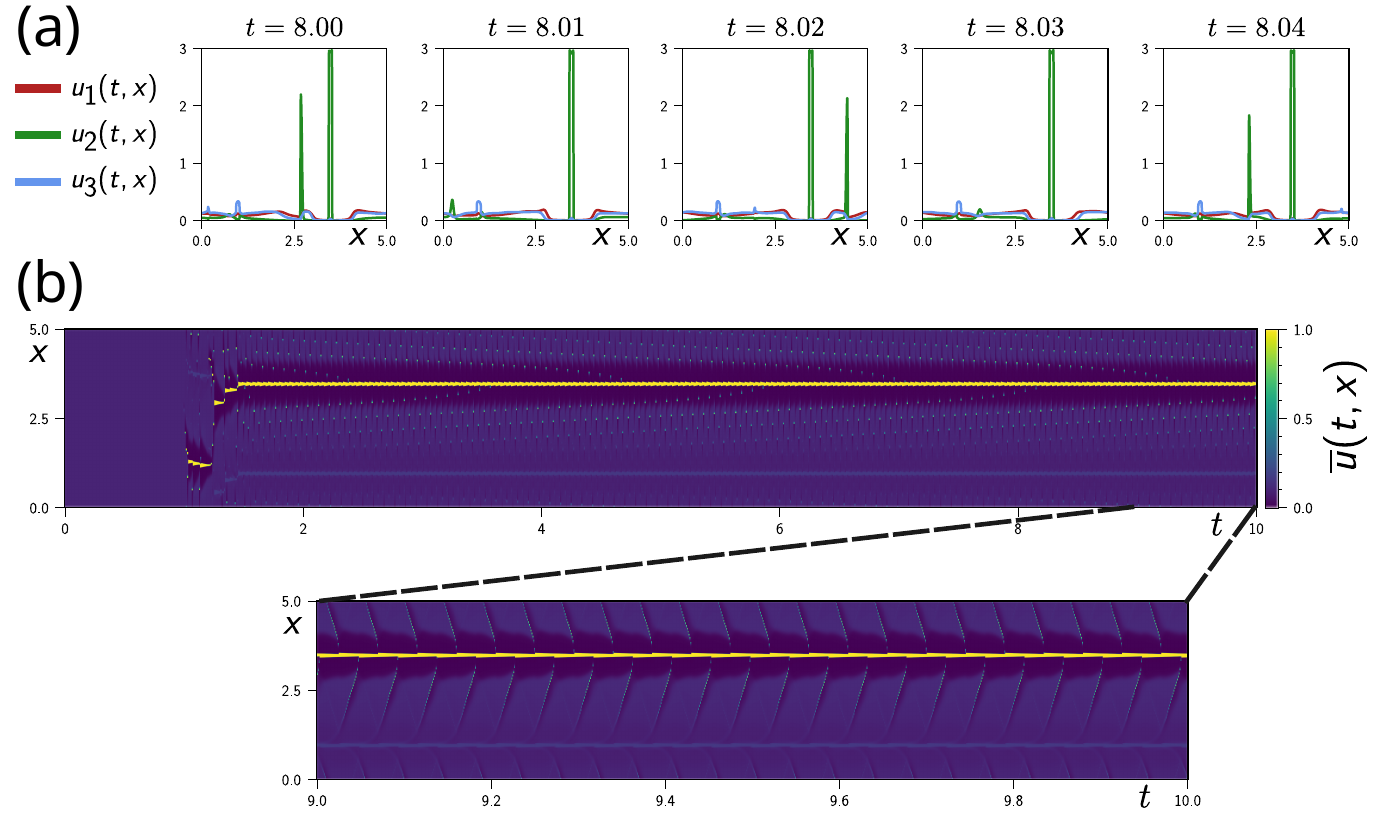}
    \caption{Numerical simulation of spatio-temporal oscillations arising from a Turing--Hopf (wave) instability. The stability matrix \eqref{eq:discrete_stability_matrix} predicts a fastest-growing mode with a growth rate that has a non-zero imaginary part. We consider $N=3$ phenotypes at an intermediate switching rate, $\beta=1$, with a sign-changing response $(B_1,B_2,B_3)=(-7.6,29,-22)$, allowing both attractive and repulsive phenotypes. Panel (a) shows the phenotype densities $u_1,u_2,u_3$ at successive times, while (b) shows a kymograph of the phenotype-averaged density $\overline{u}(t,x)$, \changes{with a magnified section highlighting the oscillations.} Parameters: $(C_1,C_2,C_3)=(1,3.9,4.2)$, $\mu=40$, $L=5$, $U=0.1$. Videos are available at \cite{My_GitHub}.
    }
    \label{fig:linear_oscillation}
\end{figure}

\begin{figure}
    \centering
    \includegraphics[width=0.87\linewidth]{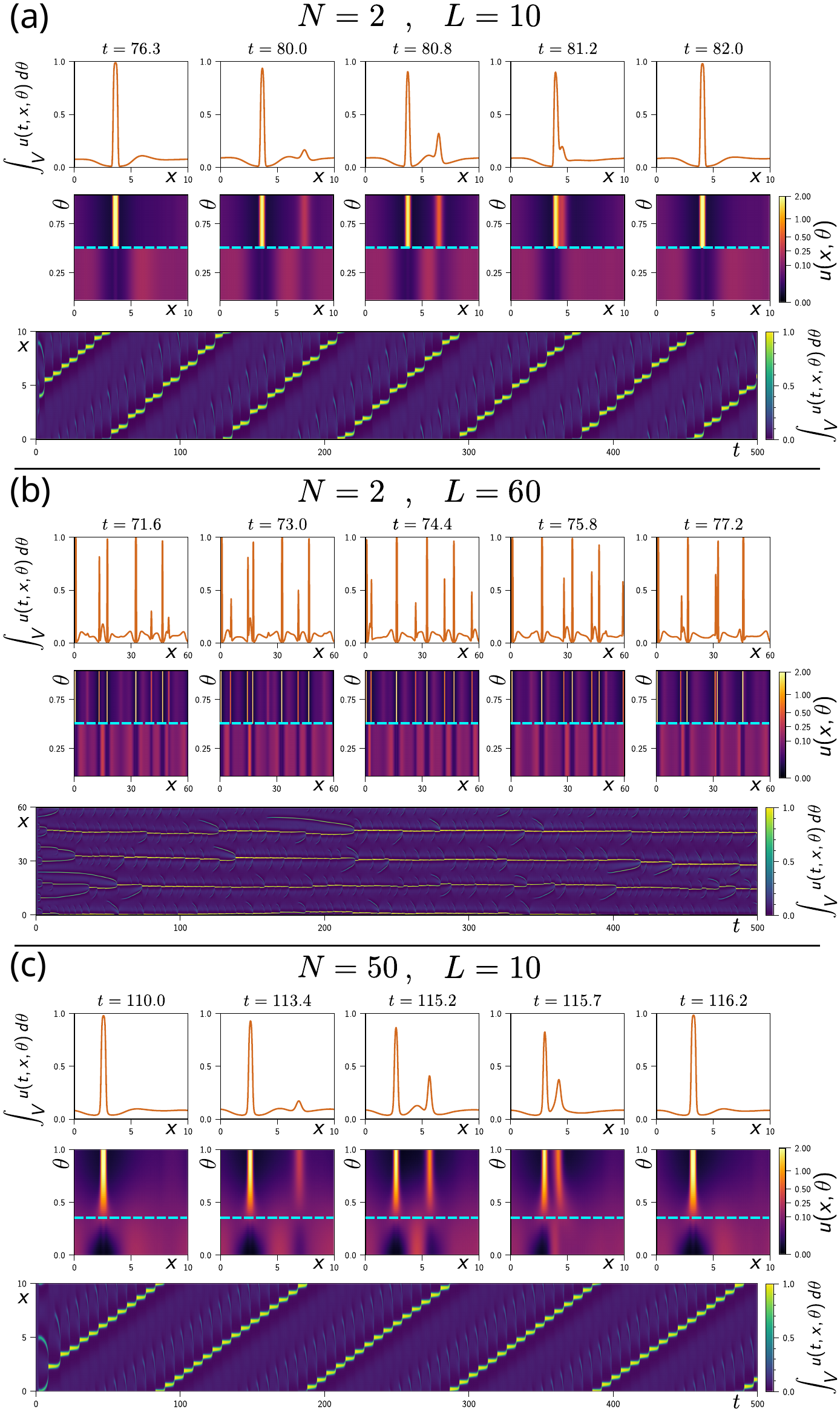}
    \caption{Numerical simulations illustrating nonlinear spatio-temporal behaviour at an intermediate phenotypic switching rate, $\beta=0.1$, with the sign-changing response function $B(\theta)=1-3\cos(\pi\theta)$, allowing both attractive and repulsive phenotypes. In (a)--(c), the upper panels show the total population density at successive times, the middle panels show the corresponding distributions in physical and phenotype space, with the cyan dashed line separating repulsive from attractive phenotypes, and the lower panels show kymographs of the total population density. Despite having only $N=2$ phenotypes in (a), compared with $N=50$ in (c), both exhibit similar stop--start travelling patterns. In contrast, the larger domain in (b), $L=60$ rather than $L=10$, produces irregular spatio-temporal dynamics resembling chaos. Parameters: $\mu=100$, $U=0.1$, $C(\theta)=1-\cos(\pi\theta)$. Videos are available at \cite{My_GitHub}.
}
    \label{fig:nonlinear_oscillations}
\end{figure}

\subsubsection{Example of linear spatio-temporal oscillations}

\changes{We now demonstrate that repulsive cross-phenotype interactions, sufficient attraction to drive pattern formation, and intermediate switching rates can together give rise to an oscillatory instability.} As a simple proof of concept, Figure~\ref{fig:linear_oscillation} considers $N=3$ phenotypes with an intermediate switching rate, $\beta=1$, and a response function $B$ containing negative values, corresponding to repulsive phenotypes. The stability matrix \eqref{eq:discrete_stability_matrix} predicts that the fastest-growing mode has a complex growth rate with positive real part and non-zero imaginary part, corresponding to a Turing--Hopf (or wave) instability. Consistent with this prediction, the numerical simulation develops a spatio-temporally oscillating pattern. This provides an example of how coupling between movement in phenotype space and movement in physical space can generate oscillations when the two occur on comparable timescales.

\subsubsection{Nonlinear spatio-temporal oscillations}

Spatio-temporal oscillations can also arise through nonlinear mechanisms that are not predicted by the linear stability analysis. Figure~\ref{fig:nonlinear_oscillations} shows examples with an intermediate phenotypic switching rate and a sign-changing response function, allowing both attractive and repulsive phenotypes. For these parameters, the linear growth rates are purely real for all $k$, and the spatio-temporal dynamics emerge only once the initially growing patterns reach sufficiently large amplitude, confirming their nonlinear origin.

The spatio-temporal behaviour arises through a nucleation--merging mechanism similar to that observed in other aggregation models \cite{painter2011_chaos_chemotaxis,painter2015nonlocal,jewell2023_dimension}. However, whilst nucleation in these \changes{cited} models is driven by proliferation, here there is no proliferation, the population is conserved, and nucleation is instead driven by switching between attractive and repulsive phenotypes. Attractive phenotypes initially form a first aggregate, driving repulsive phenotypes away to form a second aggregate out of phase. At the first aggregate, diffusion in phenotype space switches individuals from attractive to repulsive, causing them to be driven towards the second aggregate, increasing its size. At the second aggregate, phenotypic diffusion has the opposite effect, switching individuals towards attractive phenotypes. The two aggregates then attract, move together and merge, after which the cycle repeats. Thus, movement in phenotype space continually restructures the spatial pattern, which in turn determines where switching generates attractive and repulsive individuals. Such coupling appears to rely on movement in phenotype and physical space occurring on comparable timescales.

This mechanism can occur with either a small or large number of phenotypes: the $N=2$ and $N=50$ simulations in Figure~\ref{fig:nonlinear_oscillations}(a,c) exhibit very similar stop--start travelling dynamics. The character of these dynamics, however, depends strongly on the size of the physical domain. On the smaller domain, repeated nucleation and merging produce periodic stop--start motion, whereas the larger domain in Figure~\ref{fig:nonlinear_oscillations}(b) supports multiple aggregates and develops irregular dynamics resembling chaos. Indeed, applying the 0--1 test for chaos \cite{chaos_0_1_test_review,chaos_0_1_test_implementation} to the simulated time series provides numerical evidence that these dynamics are chaotic. We conjecture that increasing the domain size may lead to a period-doubling route to chaos, similar to that observed in \cite{painter2011_chaos_chemotaxis}, although we leave a systematic investigation for future work.

Interestingly, although this mechanism is nonlinear, it occurs under the same three conditions identified above as necessary for linear oscillatory instabilities under response--signal separation: intermediate phenotypic switching, repulsive interactions, and sufficient attraction to drive aggregation. This provides a further example of how coupling between movement in phenotype and physical space can generate complex spatio-temporal behaviour when the two processes occur on comparable timescales.

\section{Discussion}
\label{sec:discussion}

Phenotypic heterogeneity can fundamentally alter how populations organise collectively in space. In this work, we explored these effects using phenotype-structured nonlocal aggregation models, characterising how heterogeneity influences whether patterns form, how populations are distributed across phenotype space within them, and whether the resulting patterns are stationary or oscillatory. We considered interactions in which phenotype determines how strongly individuals produce signal and how strongly they respond to others. We analysed regimes in which phenotypic switching is slow and fast, providing useful bounds on the onset of pattern formation, before considering generic switching rates and showing how the linear growth rates interpolate monotonically between these limits.

A central result is that correlations between signal production and response strongly influence pattern formation. When individuals that produce stronger signals are also those that respond more strongly, pattern formation is promoted. Conversely, negative correlation, in which individuals specialise towards either signal production or response, inhibits pattern formation. Increasing phenotypic switching progressively dampens this effect: switching inhibits pattern formation under positive correlation but promotes it under negative correlation, with the distinction disappearing in the fast-switching limit.

Similar effects have recently been observed in a phenotype-structured Keller--Segel model\,\cite{lorenzi_painter2025phenotype_chemotaxis}. Under slow phenotypic switching, positive correlation between chemotactic sensitivity and chemoattractant secretion promotes pattern formation, whereas negative correlation inhibits it; under fast switching, the influence of these correlations again disappears. The emergence of the same qualitative behaviour in a different aggregation framework suggests that our insights may have broader generality.

We also characterised how patterns are structured in phenotype space. Under slow switching, this structure follows the response function: more responsive individuals become more concentrated at the centre of an aggregate, while less responsive individuals remain more spatially dispersed. In contrast, signal production has no effect on the phenotype structure of the pattern. Increasing phenotypic switching progressively homogenises this structure until, in the fast-switching limit, the pattern becomes uniform across phenotype space. An important consequence is that aggregation is collective: weakly interacting individuals that could not pattern in isolation can nevertheless be recruited into an aggregate through their interactions with the wider population. This extends even to individuals whose interactions are repulsive.

\changes{Another theme of this study is how phenotypic complexity can be introduced, and under what conditions this leads to spatio-temporal oscillations.} Generally in interacting systems, increasing the number of components can lead to increasingly complex eigenvalue spectra and dynamical behaviour. For example, Turing-Hopf (wave) instabilities in standard reaction--diffusion systems require three or more components \cite{Edgardo2023_wave_instabilities}, while work on complex ecological systems has demonstrated how stability depends on both the number and structure of interactions \cite{may1972_complexity_stability, allesina2012stability}. This raises an interesting challenge in contexts such as developmental patterning, where the desired outcome is often a stable spatial structure whilst populations contain substantial phenotypic complexity.

Our results provide several examples of interaction structures that \changes{can maintain stationary patterns even with substantial phenotypic complexity.} If all \changes{cross-phenotype} interactions are attractive, the leading eigenvalue is always real, so spatio-temporal oscillations cannot arise through linear instability, even in the presence of phenotypic switching. Even if some interactions are repulsive, sufficiently fast switching can homogenise the phenotypes to behave effectively as one, preventing oscillatory linear instabilities. Finally, under slow or negligible switching, response--signal separation similarly gives only real pattern-forming growth rates. This leaves the possibility for spatio-temporal oscillations under response--signal separation only when some \changes{cross-phenotype} interactions are repulsive, sufficient attraction remains to drive pattern formation, and phenotypic switching occurs on a comparable timescale to movement in physical space. Indeed, under these conditions we observe spatio-temporal behaviour arising through both linear and fully nonlinear mechanisms. \changes{In the nonlinear case, we observe a nucleation--merging mechanism similar to that seen in previous models \cite{painter2011_chaos_chemotaxis,painter2015nonlocal,jewell2023_dimension}. However, the underlying mechanism appears to be novel, driven by switching between attractive and repulsive phenotypes rather than proliferation. Both the linear and nonlinear spatio-temporal behaviour discussed above are examples of how movement in phenotype and physical space can couple to continually restructure patterns in both spaces.}

\changes{Other interaction structures can similarly restrict oscillatory behaviour; for example, \textit{detailed balance} between species interactions can give rise to gradient-flow structures\,\cite{yurij_jose_2026_bifurcation_two_species_nonlocal}}. More broadly, identifying interaction structures that constrain oscillatory dynamics may help to explain how complex heterogeneous populations maintain stationary spatial organisation.

Throughout this study, we considered both discrete and continuous phenotype formulations, each offering distinct mathematical advantages. The continuous formulation naturally lends itself to PDE and continuum methods, as illustrated by our calculation in Section~\ref{sec:sub_generic_switching_PF} for generic switching rates. The discrete formulation is more general, applying for arbitrary $N$ without requiring interactions to vary smoothly with phenotype, while also allowing the use of finite-dimensional matrix methods. For systems with many discrete phenotypes, this formulation may additionally open the door to random matrix theory, which has long been used to relate the spectra of large interaction matrices to ecological complexity and stability \cite{may1972_complexity_stability,allesina2012stability,piskovsky2024will,baron2022eigenvalues,baron2023breakdown}. Our results highlight the benefits of considering both formulations, allowing whichever mathematical tools are most \changes{straightforward} for a given problem to be used.

There are many compelling directions for future work. In Part II \cite{jewell2026phenotype_part2}, the follow-up to this work, we will move from interactions that depend on absolute phenotype to those that depend on phenotypic distance, allowing interaction strength to depend on how similar individuals are to one another. Beyond this, our treatment of phenotypic switching as \changes{homogeneous} linear diffusion is a simplification. Studies of epithelial--mesenchymal plasticity have shown that phenotypic states can differ substantially in their stability and transition properties \cite{hari_jolly2022EMT}, motivating phenotype-dependent switching rates. Nonlocal switching kernels provide another natural extension, allowing transitions between distant phenotypes \cite{cusseddu_lorenzi_2026phenotype_SKT}. Incorporating phenotype-dependent proliferation and death would be particularly relevant for modelling cancer, where proliferation, migration and phenotypic state can be strongly coupled \cite{sadhu_jolly_maini2026go_or_grow}.

\vskip6pt

\enlargethispage{20pt}

\section*{Declaration of AI use}
During the preparation of this work, the authors used OpenAI's GPT-5.5 large language model to assist with grammar, phrasing, proofreading, and aspects of coding, including debugging author-written code and assisting with the implementation of selected functions. After using this tool, the authors reviewed and edited the content and code as needed and take full responsibility for the content of the published article.

\ack{Thomas Jun Jewell is supported by funding from the Edmund J. Crampin Scholarship at Linacre College, University of Oxford; and the Engineering and Physical Sciences Research Council [EP/W524311/1].}

\appendix

\section{Numerical simulations}
\label{sec:appendix_numerics}

As discussed in Section~\ref{sec:slow_switching_insight}, discretising the continuous phenotype variable recovers the discrete formulation at the numerical level. We therefore report simulations using the continuous notation $u(x,\theta)$, $\mu$, and $\beta$, with the corresponding discrete parameters given by $\tilde{\mu}=|V|N^{-1}\mu$ and $\tilde{\beta}=|V|^{-2}N^2\beta$.

\subsection{Parameters}

Unless otherwise specified, we take the phenotype domain as $V=[0,1]$ with no-flux boundary conditions and the physical domain as $\Omega=[0,L]$, with $L=10$ and periodic boundary conditions. The phenotype domain is discretised into $N=50$ classes. Initial conditions consist of the homogeneous state $U=0.1$ perturbed independently \changes{at gridpoints} across physical and phenotype space by zero-mean Gaussian noise with standard deviation $10^{-3}$.

For all simulations presented in this study, we use the exponential spatial interaction kernel
\[
    K(s)
    =
    \frac{1}{2\xi}e^{-|s|/\xi},
    \qquad
    \xi=1.
\]
The kernel is truncated at half the domain length, $s=L/2$, so that each pair of points interacts only once across the periodic domain. We have verified that our qualitative results are robust to changing the kernel to a tophat kernel, and to variations in the interaction length scale $\xi$.

\subsection{Numerical scheme}

We discretise the physical domain using 256 grid points, giving a system of $N\times256$ ordinary differential equations. This system is integrated in time using the explicit adaptive Runge--Kutta method of order 5(4) implemented by \texttt{SciPy}. Convolutions in physical space are evaluated using Fast Fourier Transforms (FFT). The full code and further details are available at \cite{My_GitHub}.

\subsection{Dynamic Mode Decomposition}
\label{sec:appendix_dmd}

To independently validate the analytical dispersion relations, we infer linear growth rates directly from the numerical simulations of the full nonlinear system using Dynamic Mode Decomposition (DMD) \cite{DMD}. DMD approximates the linear evolution between successive snapshots of a dynamical system and extracts its eigenmodes and associated eigenvalues directly from time-series data.

We first subtract the homogeneous state and Fourier transform the simulation data in physical space. \changes{Exact DMD is then applied independently to the dynamics associated with each spatial Fourier mode $k$, yielding a set of discrete-time eigenvalues $\rho_j(k)$, indexed by $j$. For snapshots separated by $\Delta t$, the corresponding continuous-time exponential growth rates are}
\begin{equation}
    \lambda_j(k)
    =
    \frac{\log\rho_j(k)}{\Delta t}.
\end{equation}
The numerically inferred dispersion relation is then given by the largest real growth rate at each wavenumber, 
\begin{equation}
    \lambda(k)
    =
    \max_j \operatorname{Re}\!\left[\lambda_j(k)\right].
\end{equation}
DMD is applied to snapshots from the early stages of each simulation, when perturbations remain small and the dynamics are approximately linear, while excluding the earliest times associated with relaxation of the random initial condition. In practice, we use the time window $t\in[0.1,0.1+6N\Delta t]$ with $\Delta t=10^{-3}$ \changes{(i.e. typically $t\in[0.1,0.4]$)}. The analysis requires no knowledge of the model parameters beyond the physical domain geometry and provides an independent numerical test of the analytical dispersion relations. We implement DMD using the \texttt{PyDMD} package.

\section{Solving the eigenvalue problem for generic switching}
\subsection{Deriving the Green's function}
\label{sec:appendix_greens_function}

For generic phenotypic switching, $0<\beta<\infty$, solving the phenotype eigenvalue problem \eqref{eq:continuous_eigenproblem} requires solving the second-order ODE \eqref{eq:generic_switching_continuous_ode}. In the main text, we do this by introducing the Green's function, $G_\lambda(\theta,\theta')$, defined in \eqref{eq:greens_function_definition}. Here, we derive this Green's function.

Let 
\[
\kappa=\sqrt{\frac{k^2+\lambda(k)}{\beta}}.
\]
Since it is not known a priori whether $\lambda(k)$ is real or satisfies $\lambda(k)+k^2>0$, we do not assume that $\kappa$ is real. Accordingly, the hyperbolic functions below are interpreted in the usual way for complex arguments. Equation \eqref{eq:greens_function_definition} implies that, for $\theta\neq\theta'$, the Green's function satisfies
\[
\frac{\partial^2 G_\lambda(\theta,\theta')}{\partial\theta^2}
-
\kappa^2 G_\lambda(\theta,\theta')=0.
\]
Assuming that $V=[\,0,\,|V|\,]$, the general solution to the above ODE for $\theta<\theta'$ and $\theta>\theta'$ is
\[
G_\lambda(\theta,\theta')
=
\begin{cases}
a(\theta')\cosh(\kappa\theta)+b(\theta')\sinh(\kappa\theta),
& \theta<\theta',\\[2mm]
c(\theta')\cosh\!\bigl(\kappa(|V|-\theta)\bigr)
+
d(\theta')\sinh\!\bigl(\kappa(|V|-\theta)\bigr),
& \theta>\theta'.
\end{cases}
\]
The Neumann boundary conditions imply $b(\theta')=d(\theta')=0$. Continuity at $\theta=\theta'$ requires that
\[
a(\theta')\cosh(\kappa\theta')
=
c(\theta')\cosh\!\bigl(\kappa(|V|-\theta')\bigr).
\]
Finally, integrating the defining equation \eqref{eq:greens_function_definition} across a small neighbourhood of $\theta'$ yields the jump condition
\[
\left.\frac{\partial G_\lambda}{\partial\theta}\right|_{\theta=\theta'^+}
-
\left.\frac{\partial G_\lambda}{\partial\theta}\right|_{\theta=\theta'^-}
=
-\frac{1}{\beta}.
\]
Provided $\kappa\sinh(\kappa|V|)\neq0$, combining these conditions gives
\[
a(\theta')=
\frac{\cosh\!\bigl(\kappa(|V|-\theta')\bigr)}{\beta\kappa\sinh(\kappa|V|)},
\qquad
c(\theta')=
\frac{\cosh(\kappa\theta')}{\beta\kappa\sinh(\kappa|V|)}.
\]
The Green's function is thus given by
\begin{equation}
\begin{split}
    G_\lambda(\theta,\theta')
=
\begin{cases}
\dfrac{\cosh(\kappa\theta)\cosh\!\bigl(\kappa(|V|-\theta')\bigr)}
{\beta\kappa\sinh(\kappa|V|)},
& \theta<\theta',\\[3mm]
\dfrac{\cosh(\kappa\theta')\cosh\!\bigl(\kappa(|V|-\theta)\bigr)}
{\beta\kappa\sinh(\kappa|V|)},
& \theta>\theta'.
\end{cases}
\end{split}
\label{eq:greens_function_appendix}
\end{equation}

\subsection{Calculating Green's function integrals}
\label{sec:appendix_integral_calculation}

\subsubsection{Positively correlated response and signal functions}
\label{sec:appendix_integral_positive_correlation}

With the choice
\[
B(\theta)=C(\theta)=1-\cos(\pi\theta),
\qquad
\theta\in V=[0,1],
\]
the Green's function integral \eqref{eq:I_lambda} becomes
\[
\mathcal{I}_\lambda[B,C]
=\mathcal{I}_\lambda[B,B]
=
\int_0^1\int_0^1
B(\theta)\,G_\lambda(\theta,\theta')\,B(\theta')
\,\mathrm{d}\theta'\,\mathrm{d}\theta.
\]
Since the Green's function \eqref{eq:greens_function_appendix} is symmetric under interchange of $\theta$ and $\theta'$, the integrand is also symmetric, and so the double integral may be written as
\[
\mathcal{I}_\lambda[B,C]
=
\frac{2}{\beta\kappa\sinh(\kappa)}
\int_0^1
\int_0^\theta
\left(1-\cos(\pi\theta)\right)
\left(1-\cos(\pi\theta')\right)
\cosh(\kappa\theta')
\cosh\!\bigl(\kappa(1-\theta)\bigr)
\,\mathrm{d}\theta'
\,\mathrm{d}\theta.
\]
Evaluating the inner integral and then the outer integral using standard trigonometric and hyperbolic integral identities gives
\[
\mathcal{I}_\lambda[B,C]
=
\frac{1}{\beta\kappa^2}
+
\frac{1}{2\beta(\kappa^2+\pi^2)}
\equiv
\frac{1}{\bigl(k^2+\lambda(k)\bigr)}
+
\frac{1}{2\bigl(k^2+\lambda(k)+\beta\pi^2\bigr)}.
\]

\subsubsection{Negatively correlated response and signal functions}
\label{sec:appendix_integral_negative_correlation}

We next consider the negatively correlated choice
\[
B(\theta)=1-\cos(\pi\theta),
\qquad
C(\theta)=1+\cos(\pi\theta),
\qquad
\theta\in V=[0,1].
\]
Noting that
\[
C(\theta)=2-B(\theta),
\]
the Green's function integral \eqref{eq:I_lambda} can be written, using linearity, as
\[
\mathcal{I}_\lambda[B,C]
=
\mathcal{I}_\lambda[B,2]
-
\mathcal{I}_\lambda[B,B].
\]
Since $\mathcal{I}_\lambda[B,B]$ was evaluated in Appendix~\ref{sec:appendix_integral_calculation}\ref{sec:appendix_integral_positive_correlation}, it remains only to calculate
\[
\mathcal{I}_\lambda[B,2]
=
2\int_0^1\int_0^1
G_\lambda(\theta,\theta')\,B(\theta')
\,\mathrm{d}\theta'\,\mathrm{d}\theta.
\]
From the Green's function \eqref{eq:greens_function_appendix},
\[
\begin{split}
\int_0^1
G_\lambda(\theta,\theta')
\,\mathrm{d}\theta
={}&
\frac{\cosh\!\bigl(\kappa(1-\theta')\bigr)}
{\beta\kappa\sinh(\kappa)}
\int_0^{\theta'}
\cosh(\kappa\theta)\,\mathrm{d}\theta
\\
&
+
\frac{\cosh(\kappa\theta')}
{\beta\kappa\sinh(\kappa)}
\int_{\theta'}^1
\cosh\!\bigl(\kappa(1-\theta)\bigr)\,\mathrm{d}\theta.
\end{split}
\]
Evaluating the two integrals gives
\[
\int_0^1
G_\lambda(\theta,\theta')
\,\mathrm{d}\theta
=
\frac{
\cosh\!\bigl(\kappa(1-\theta')\bigr)\sinh(\kappa\theta')
+
\cosh(\kappa\theta')\sinh\!\bigl(\kappa(1-\theta')\bigr)
}
{\beta\kappa^2\sinh(\kappa)}=\frac{1}{\beta\kappa^2},
\]
where for the second equality, we used hyperbolic addition identities.

Hence,
\[
\mathcal{I}_\lambda[B,2]
=
\frac{2}{\beta\kappa^2}
\int_0^1
\left(
1
-
\cos(\pi\theta')
\right)
\,\mathrm{d}\theta'
=
\frac{2}{\beta\kappa^2}.
\]
Together with the result for $\mathcal{I}_\lambda[B,B]$ from Appendix~\ref{sec:appendix_integral_calculation}\ref{sec:appendix_integral_positive_correlation}, this gives
\begin{equation*}
\begin{split}
    \mathcal{I}_\lambda[B,C]
&=\frac{2}{\beta\kappa^2}-\left(\frac{1}{\beta\kappa^2}
+
\frac{1}{2\beta(\kappa^2+\pi^2)}\right)=\frac{1}{\beta\kappa^2}
-
\frac{1}{2\beta(\kappa^2+\pi^2)} \\
&\equiv
\frac{1}{\bigl(k^2+\lambda(k)\bigr)}
-
\frac{1}{2\bigl(k^2+\lambda(k)+\beta\pi^2\bigr)}.
\end{split}
\end{equation*}

\vskip2pc

\bibliographystyle{RS}
\bibliography{references}

\end{document}